\documentclass[11pt,a4paper]{article}
\usepackage[T1]{fontenc}
\usepackage[utf8]{inputenc}
\usepackage[english]{babel}
\usepackage{amsmath,amssymb,amsthm,mathtools}
\usepackage{newtxtext,newtxmath}
\usepackage{mathrsfs}
\usepackage[a4paper,top=25mm,bottom=25mm,left=26mm,right=26mm]{geometry}
\usepackage{microtype,enumitem,booktabs,longtable,array,etoolbox}
\usepackage{fancyhdr}
\usepackage{needspace}
\usepackage{xcolor}

\usepackage[hidelinks]{hyperref}
\usepackage{bookmark}
\InputIfFileExists{ursfs.fd}{}{}
\DeclareFontShape{U}{rsfs}{m}{n}{<-5> rsfs5 <5> <5.5> <6> <6.1> rsfs5 <6.6> <7> rsfs7 <7.9> <8> <8.8> <9> <9.4> <10> <10.5> <10.95> <12> <12.5> <14.4> <17.28> <20.74> <24.88> rsfs10}{}

\numberwithin{equation}{section}
\newtheorem{theorem}{Theorem}[section]
\newtheorem{lemma}[theorem]{Lemma}
\newtheorem{proposition}[theorem]{Proposition}
\newtheorem{corollary}[theorem]{Corollary}
\theoremstyle{definition}
\newtheorem{definition}[theorem]{Definition}
\newtheorem{example}[theorem]{Example}
\theoremstyle{remark}
\newtheorem{remark}[theorem]{Remark}

\newcommand{\R}{\mathbb R}
\newcommand{\C}{\mathbb C}
\newcommand{\N}{\mathbb N}
\newcommand{\Nn}{\mathbb N_0}
\newcommand{\PP}{\mathcal P}
\newcommand{\ER}{\mathcal E_{\R}}
\newcommand{\QQ}{\mathcal Q}
\newcommand{\DD}{\mathcal D}
\newcommand{\Rec}{\mathcal R}

\newcommand{\hc}{\mathbf h}
\newcommand{\ct}{\mathtt c}
\newcommand{\st}{\mathtt s}

\DeclareMathOperator{\Dom}{Dom}

\setlist[enumerate]{topsep=4pt,itemsep=3pt,parsep=0pt,leftmargin=*}
\allowdisplaybreaks[2]

\newcommand{\proofstep}[1]{\par\smallskip\noindent\textbf{#1}\quad}
\newcommand{\cm}{c_{\mathrm{mod}}}
\newcommand{\cd}{c_{\mathrm{dir}}}

\hypersetup{
    pdftitle={Sharp Query Bounds and Optimal Faithful Simulation under Smooth Changes of Variables},
    pdfauthor={Rodrigo Alberto Castro Marín},
    pdfsubject={Exact representations and optimal faithful simulation of derivative-query procedures},
    bookmarksnumbered=true
}

\title{
    \textbf{Sharp Query Bounds and Optimal Faithful Simulation}\\[5pt]
    \large under Smooth Changes of Variables
}

\author{
    Rodrigo Alberto Castro Marín\\[3pt]
    \small Institute of Mathematics, Universidad de Valparaíso, Chile\\[3pt]
    \small \href{mailto:rodrigo.castrom@uv.cl}
    {\texttt{rodrigo.castrom@uv.cl}}\\[3pt]
    \small ORCID:
    \href{https://orcid.org/0000-0003-0010-7775}
    {0000-0003-0010-7775}
}

\date{}

\begin{document}
\maketitle

\begin{abstract}
We consider deterministic procedures that acquire information about an unknown real-valued function through scalar queries to its values and derivatives. We use two descriptions of computation, one in terms of states and transitions and another in terms of rules depending on the history of responses, allowing partially defined rules. We establish representations in both directions that exactly preserve the domains of valid computations within a finite budget, their outputs, records, and counts by derivative order. These representations make it possible to study the cost of reproducing derivative information under changes of variables. For a composite function $f\circ\chi$, with $\chi$ a known smooth diffeomorphism, we prove that the minimum number of queries to $f$ needed to reconstruct a derivative response equals the number of nonzero coefficients in the composition formula, even among adaptive procedures. We extend this characterization to faithful simulation of complete procedures and obtain the exact cost from the union of the active data in their records. For successive changes of variables, we determine the costs of reconstructing intermediate responses and of computing only the final derivative, and characterize the savings due to cancellation of data. Finally, upon composing a smooth change with its inverse, we prove that the direct cost is one, whereas the intermediate reconstruction cost remains equal to one if all higher derivatives of the change vanish at the point and tends to infinity with the order otherwise. In particular, affine changes are exactly those that admit a uniform bound on this cost over all points and orders.
\end{abstract}

\clearpage

\par\medskip
\needspace{8\baselineskip}

\section{Introduction}\label{sec:introduccion}

In a numerical procedure, we may evaluate a function and some of its derivatives at points determined from previously acquired information. The value produced by the procedure does not, by itself, identify the queries made or the conditions under which it terminates. Indeed, an evaluation may have no response when the node does not belong to the domain of the function, and a computational rule may be undefined for the data received. Therefore, if we wish to study how much information a procedure uses, we must distinguish the mapping to be computed, the sequence of queries performed, and the internal operations leading to the output.

The study of algorithms through acquired information has precedents in information-based complexity. Traub and Wo\'zniakowski~\cite{traubwozniakowski} develop a general framework for analyzing algorithms and information, and Werschulz~\cite{werschulz} provides an exposition of this theory. The possibility of selecting data according to previous responses is part of that approach. Likewise, Hausmann and Korte~\cite{hausmannkorte} study oracle algorithms for fixed-point problems. In another setting, Berry and Curien~\cite{berrycurien} relate concrete and abstract descriptions of sequential algorithms, while Blass and Gurevich~\cite{blassgurevich} formalize algorithms interacting with their environment through queries and responses. Buhrman and de Wolf~\cite{buhrman} present, from another perspective, a survey of query complexity and decision trees. These precedents show that neither adaptive acquisition of information nor the description of an algorithm by states constitutes, by itself, a new notion. Their objects and hypotheses are not identified here with those we shall consider.

Fix a function $f\in C^\infty(I_f)$, where $I_f$ is an open interval of the real line that is not supplied to the procedure. A query is determined by a pair $(z,r)$, with $z\in\R$ and $r\in\Nn$, and returns the unique number $f^{(r)}(z)$ when $z\in I_f$. The rules selecting queries or producing outputs are fixed independently of $f$ and may be partial; we require neither continuity nor computability. A query outside the interval provides neither a response nor a signal that can be used in the computation. Internal operations are not counted as evaluations. Within this framework, we shall study two descriptions: a method whose decisions depend on the initial datum and the response history, and a procedure whose states distinguish internal transitions, queries, and outputs. We do not intend thereby to introduce a new general notion of oracle algorithm. Our interest is to establish exactly which properties can be preserved when passing from one description to the other.

For each procedure $\mathscr C$ and each integer budget $n\ge1$, we construct a query method $T_{\mathscr C,n}$ whose domain coincides with the inputs on which $\mathscr C$ has a valid computation using at most $n$ evaluations; this equality is established in~\eqref{eq:exactdomain}. We also prove that the output, the ordered record of queries and responses, and the number of evaluations of each order are preserved. Conversely, we represent every query method with finite budget by a procedure preserving these properties. The correspondence concerns observable behavior and does not require identifying the internal states of the two objects. We also prove that repeated responses can be stored during a computation, so that each distinct node--order pair is acquired only once without changing the domain or the output. Furthermore, we represent staged implementations and determine their evaluation counts. These constructions fix the information properties that we shall need to preserve in subsequent applications.

Consider now a known smooth diffeomorphism $\chi:J\to K$ and an unknown function $f$ defined on an open subinterval of $K$. If a procedure operates on $g=f\circ\chi$, its queries to $g$ must be reconstructed from responses obtained from $f$. The chain rule provides a linear combination of derivatives of $f$ for each derivative of $g$; the corresponding identity is given in~\eqref{eq:chi-chain}. The composition formulas and Fa\`a di Bruno coefficients are classical. Scholz and Berger~\cite{scholzberger} study their computation using Bell polynomials, while Naumann~\cite{naumann} analyzes a problem concerning the arithmetic complexity of the chain rule. Here we consider a different cost: the number of indispensable scalar queries when $f$ is known only through the responses acquired.

We prove that every derivative datum whose coefficient in the composition formula is nonzero must be queried in order to reconstruct the requested response exactly. This necessity remains even if the procedure can adaptively choose other nodes and orders. To prove it, we use Hermite interpolation: if an active datum has not been queried, we construct a polynomial perturbation that preserves all acquired responses and changes the derivative whose reconstruction is required. In this way, we obtain a lower bound matching the number of queries in the direct construction. We then extend the result to complete procedures. The simulation preserves the domain, output, and virtual record of queries to $g$, and attains the minimum cost given by the cardinality of the union of the active pairs appearing in that record. The conclusion concerns faithful reconstruction of the responses, not minimization over all procedures producing only the same final output.

The preceding distinction has consequences when changes of variables are applied successively. We may reconstruct the derivative responses at each stage or first simplify the composition to obtain only the final derivative. We determine the minima for both tasks from the supports of the intermediate expressions and the resulting expression. We prove that savings occur if and only if some datum needed to reconstruct an intermediate response disappears through complete cancellation of its coefficient in the final expression. The matrix description of composition and functional inversion has precedents in Jabotinsky~\cite{jabotinsky}, Aschenbrenner~\cite{aschenbrenner}, and Johnston and Prochno~\cite{johnston}; we do not claim novelty for those identities. What we determine is the exact cost of acquiring the information needed for the two specified tasks.

In particular, if we compose the exponential change with the logarithm, the derivative of order $r$ of the resulting function is obtained with one query, whereas faithful reconstruction of the intermediate responses requires $r$ queries. We prove that this difference is not confined to that example. For a smooth diffeomorphism $\chi$ and its inverse, we fix a point $b$ and prove that the intermediate reconstruction cost is one at every order if $\chi^{(m)}(b)=0$ for each $m\ge2$. If any of these derivatives does not vanish, that cost tends to infinity as the order increases, even though the overall composition is the identity. The proof does not assume analyticity. Consequently, affine changes are exactly those for which the modular cost remains uniformly bounded over all points and orders.

The contributions of this work therefore develop in three directions. First, we establish exact representations for procedures with derivative queries and partial rules, preserving domains, outputs, records, and counts under finite budgets. Second, we use these representations to construct faithful simulations under smooth changes and prove the necessity of every acquisition involved in their responses. Finally, we characterize reconstruction costs under successive changes and obtain the dichotomy and affine characterization described above. The precise formulation of domains and records makes it possible to distinguish these statements from results about arithmetic work, execution times, or the cost of procedures that need only reproduce an output. The lower bounds use the class of smooth functions, in which finite derivative data can be prescribed freely; we do not transfer them without justification to function classes subject to additional restrictions. The query framework is also developed, with other objectives related to order barriers, in the preprint deposited in Zenodo~\cite{castromarinzenodo}.

We begin by defining query methods and computational procedures. We then establish representations in both directions, response reuse, and the staged construction. Finally, we study optimal simulation under changes of variables, successive compositions, and the consequences of reconstruction costs.

\par\medskip
\needspace{8\baselineskip}

\section{Query Model}\label{sec:consultas}

Let us fix the conventions needed to define queries and their executions.

\subsection{Conventions and Basic Results}

We write \[
\N=\{1,2,\ldots\},
\qquad
\Nn=\{0,1,2,\ldots\}.
\]

We denote the empty list by $()$. We adopt the convention that an empty sum is zero and an empty product is one. For every function $f$, we write $f^{(0)}:=f$.

Let $I\subseteq\R$ be an open interval. We use the notations $C^s(I)$ and $C^\infty(I)$ for functions with continuous derivatives up to order $s$ and of all orders, respectively.

We say that a function $f:I\longrightarrow\R$ is \emph{real analytic} if, for every $a\in I$, there exist $\rho>0$ and real coefficients $(b_j)_{j\ge0}$ such that \[
(a-\rho,a+\rho)\subseteq I
\] and \[
f(a+t)=\sum_{j=0}^{\infty}b_jt^j,
\qquad |t|<\rho.
\]

A function $F:\C\longrightarrow\C$ is called \emph{entire} if it admits a power-series expansion in a neighborhood of every point of $\C$.

Finally, we define \[
\boxed{
\ER=
\left\{
F:\C\longrightarrow\C:
F\text{ is entire and }
F(\R)\subseteq\R
\right\}.
}
\]

\begin{lemma}[Reality of the derivatives]\label{lem:real}
If $F\in\ER$, then $F^{(r)}(x)\in\R$ for every $x\in\R$ and $r\in\Nn$. The restriction $F|_\R$ is real analytic, belongs to $C^\infty(\R)$, and its real derivatives coincide with the restrictions of the complex derivatives.
\end{lemma}

\begin{proof}
Let $F\in\ER$. Since $F(\R)\subseteq\R$, we have $F(x)\in\R$ for every $x\in\R$. We prove by induction that $F^{(r)}(x)\in\R$ for every $r\in\Nn$.

For $r=0$, the statement is immediate. Suppose that $F^{(r)}(x)\in\R$ for every $x\in\R$. Since $F^{(r)}$ is entire, we have \[
F^{(r+1)}(x)
=\lim_{\substack{t\to0\\t\in\R}}
\frac{F^{(r)}(x+t)-F^{(r)}(x)}{t}.
\] Note that, by the induction hypothesis, the preceding quotient is real for every $t\in\R\setminus\{0\}$. Therefore, its limit also belongs to $\R$, that is, $F^{(r+1)}(x)\in\R$.

Moreover, this limit coincides with the real derivative of the restriction of $F^{(r)}$ to $\R$. Thus, by induction, we obtain \[
F^{(r)}(x)\in\R,\qquad
(F|_\R)^{(r)}(x)=F^{(r)}(x)
\] for every $x\in\R$ and $r\in\Nn$. Since the derivatives of an entire function are continuous, we conclude that $F|_\R\in C^\infty(\R)$.

Finally, let $a\in\R$. Since $F$ is entire, there exists $\rho>0$ for which \[
F(a+z)=\sum_{j=0}^{\infty}
\frac{F^{(j)}(a)}{j!}z^j,
\qquad |z|<\rho.
\] Restricting this equality to $z=t\in\R$, all coefficients are real by what has already been proved. We conclude that $F|_\R$ is real analytic.
\end{proof}

When a function has several roots and one of them is chosen, we call it a \emph{distinguished root}.

\subsection{Rules, Queries, and Responses}

\begin{definition}[Total and partial rules]\label{def:rule}
Let $X,Y$ be sets. A \emph{data rule} from $X$ to $Y$ is a mapping $\Phi:D_\Phi\to Y$, where $D_\Phi\subseteq X$. We say that $\Phi$ is \emph{total} if $D_\Phi=X$ and \emph{properly partial} if $D_\Phi\ne X$. In what follows, the term \emph{partial rule} includes both cases.

We do not require $D_\Phi$ to be open or $\Phi$ to be continuous, whenever these notions apply. Nor do we assume that there exists an algorithm for deciding membership in $D_\Phi$.
\end{definition}

\begin{definition}[Query and response]\label{def:query}
Consider the set $\QQ=\R\times\Nn$. An element $q=(z,r)\in\QQ$ is called a \emph{query}, where $z$ is the \emph{node} and $r$ the \emph{derivative order}.

Let $f\in C^\infty(I_f)$, where $I_f\subseteq\R$ is an open interval. We define \[
\QQ_f=I_f\times\Nn,\qquad
E_f:\QQ_f\to\R,\qquad
E_f(z,r)=f^{(r)}(z).
\] For each $q=(z,r)\in\QQ_f$, we call the value $E_f(q)$ the \emph{response} and the triple $(z,r,E_f(z,r))$ an \emph{answered query}. Note that the response is defined only for queries belonging to $\QQ_f$.

Finally, if $F\in\ER$, by Lemma~\ref{lem:real}, its restriction to $\R$ belongs to $C^\infty(\R)$ and, therefore, $\QQ_F=\QQ$.
\end{definition}

Note that every query performed is counted once, even if the same pair $(z,r)$ is requested several times. By contrast, reusing a response already obtained does not constitute a new query. Therefore, this count does not include arithmetic operations or the cost of computing derivatives.

\subsection{Data and Query or Output Rules}
For each $k\in\Nn$, consider the space $H_k=\R^{k+1}$, whose elements we write as \[
\hc_k=(x,\eta_1,\ldots,\eta_k)
\] and call \emph{level-$k$ data vectors}. For $0\le i\le k$, we define the \emph{level-$i$ prefix} by \[
\pi_i\hc_k=(x,\eta_1,\ldots,\eta_i)\in H_i.
\] In particular, we identify $\hc_0=(x)$ with $x$ whenever no ambiguity arises.

Fix two distinct tags $\ct,\st$ and define the disjoint sets \[
\mathcal A_c=\{\ct\}\times\QQ,\qquad
\mathcal A_s=\{\st\}\times\R,\qquad
\mathcal A=\mathcal A_c\cup\mathcal A_s.
\] The elements $(\ct,q)\in\mathcal A_c$ and $(\st,a)\in\mathcal A_s$ are called \emph{query decisions} and \emph{output decisions}, respectively. In the first case, $q\in\QQ$ determines the query and, in the second, $a\in\R$ determines the output value.

\begin{definition}[Method with budget $d$]\label{def:method}
Let $d\in\N$. For each $0\le k\le d$, consider $D_k\subseteq H_k$ and the mappings \[
\Phi_k:D_k\to\mathcal A,\quad 0\le k<d,
\qquad
\Phi_d:D_d\to\mathcal A_s.
\] We call the tuple \[
T=\bigl(d,(D_k)_{k=0}^{d},(\Phi_k)_{k=0}^{d}\bigr).
\] a \emph{deterministic, causal, memoryless method with budget $d$}. The domains $D_k$ and mappings $\Phi_k$ are fixed independently of the function $f$ to which the method is applied.

The method is \emph{deterministic} because each rule $\Phi_k$ assigns a unique decision to every element of its domain, and it is \emph{causal} because the decision at level $k$ depends only on the data $(x,\eta_1,\ldots,\eta_k)$. We say that it is \emph{memoryless} because each new iteration starts only from the datum $x$, without using information from previous iterations.

Finally, we say that $T$ is \emph{total} if $D_k=H_k$ for every $0\le k\le d$, and \emph{has partially defined rules} otherwise.
\end{definition}

Note that the rules of the method may use constants fixed independently of $f$, but they have no access to its expression, coefficients, root, or any other information about $f$ that has not been acquired through queries. Moreover, we do not require these rules to be computable.

\subsection{Executions Defined by Equations}

\begin{definition}[Admissible prefix]\label{def:prefix}
Let $T$ be a method as in Definition~\ref{def:method}, let $f\in C^\infty(I_f)$ and $x\in\R$. For $0\le k\le d$, consider a list \[
\Rec_k=((z_j,r_j,\eta_j))_{j=1}^{k}.
\] We say that $\Rec_k$ is an \emph{admissible prefix of length $k$} if, on defining $q_j=(z_j,r_j)$ and $\hc_i=(x,\eta_1,\ldots,\eta_i)$, we have \begin{equation}\label{eq:prefix}
\hc_i\in D_i,\qquad
\Phi_i(\hc_i)=(\ct,q_{i+1}),\qquad
q_{i+1}\in\QQ_f,\qquad
\eta_{i+1}=E_f(q_{i+1}),
\quad 0\le i<k.
\end{equation} For $k=0$, the empty list $\Rec_0=()$ is regarded as an admissible prefix of length zero.
\end{definition}

\begin{definition}[Valid execution, output, and termination]
\label{def:execution}
Let $T$ be a method as in Definition~\ref{def:method}. We say that a pair $(\Rec_k,a)$ is a \emph{valid execution} of $T$ on $f$ starting from $x$ if $\Rec_k$ is an admissible prefix of length $k$ and, in addition, \begin{equation}\label{eq:exit}
\hc_k\in D_k,\qquad \Phi_k(\hc_k)=(\st,a).
\end{equation} In this case, $\Rec_k$ is called the \emph{record} of the execution, $a$ is its \emph{output}, and $k$ is the \emph{number of queries performed}. We say that the method terminates after $k$ queries with output $a$ when \eqref{eq:prefix} and \eqref{eq:exit} hold.

If no valid execution exists on $(f,x)$, we say that the method is \emph{undefined} at $(f,x)$. Note that if a query $q\notin\QQ_f$, its response $E_f(q)$ is undefined, and this circumstance provides no additional information that the rules of the method could use.
\end{definition}

\begin{lemma}[Uniqueness of prefixes and executions]
\label{lem:unique}
Fix $T$, $f$, and $x$. Then there is at most one admissible prefix of each length and at most one valid execution.
\end{lemma}

\begin{proof}
Suppose that $\Rec_k$ and $\Rec_k'$ are two admissible prefixes of length $k$. We show by induction that they coincide.

For $k=0$, the statement is immediate, since both prefixes are the empty list. Suppose that the data coincide up to level $j<k$, that is, $\hc_j=\hc_j'$. Since the rules $\Phi_j$ are mappings and both prefixes are admissible, we have \[
\Phi_j(\hc_j)=\Phi_j(\hc_j')
\quad\Longrightarrow\quad
q_{j+1}=q_{j+1}'.
\] Therefore, \[
\eta_{j+1}=E_f(q_{j+1})
=E_f(q_{j+1}')=\eta_{j+1}',
\] whence $\hc_{j+1}=\hc_{j+1}'$. As the initial data coincide, $\hc_0=\hc_0'=(x)$, we conclude by induction that $\Rec_k=\Rec_k'$.

We now prove uniqueness of the execution. Suppose that $(\Rec_k,a)$ and $(\Rec_\ell',a')$ are two valid executions. If $k<\ell$, then, by the preceding argument, the first $k$ data of both executions coincide, that is, $\hc_k=\hc_k'$. However, by the definition of a valid execution, \[
\Phi_k(\hc_k)=(\st,a)\in\mathcal A_s,
\qquad
\Phi_k(\hc_k')=(\ct,q_{k+1}')\in\mathcal A_c,
\] which contradicts $\mathcal A_s\cap\mathcal A_c=\varnothing$. Exchanging the two executions also rules out $\ell<k$. Thus, $k=\ell$ and the prefixes coincide; uniqueness of the decision $\Phi_k$ gives $a=a'$. We conclude that the valid execution is unique, if it exists.
\end{proof}

By Lemma~\ref{lem:unique}, we may define the set \[
\Omega_{T,f}
=\{x\in\R:\text{there exists a valid execution of $T$
on $(f,x)$}\}.
\] For each $x\in\Omega_{T,f}$, if $(\Rec_k,a)$ is the corresponding valid execution, we define \[
T_f:\Omega_{T,f}\to\R,\qquad T_f(x)=a,
\qquad \nu_{T,f}(x)=k.
\] Note that these mappings are well defined by uniqueness of the execution and that $0\le \nu_{T,f}(x)\le d, \qquad x\in\Omega_{T,f}.$

We call each application of $T_f$ an \emph{iteration}. If, starting from $x_0$, a sequence $(x_n)$ is defined by \[
x_{n+1}=T_f(x_n),\qquad n\in\Nn,
\] and this relation is defined for every $n$, we call that sequence a \emph{trajectory}.

\begin{proposition}[Existence for total methods]
\label{prop:total}
If $T$ is a total method and $F\in\ER$, then $\Omega_{T,F}=\R$.
\end{proposition}

\begin{proof}
Let $x\in\R$. Since $T$ is total, we have $D_k=H_k$ for every $0\le k\le d$. In particular, $\hc_0=(x)\in D_0$ and the empty list $\Rec_0=()$ is an admissible prefix.

Suppose that the method has not terminated after $k<d$ queries and that an admissible prefix $\Rec_k$ has been obtained. Since $\hc_k\in D_k$, we have $\Phi_k(\hc_k)\in\mathcal A$. Therefore, there are two cases: if $\Phi_k(\hc_k)=(\st,a)$, then $(\Rec_k,a)$ is a valid execution and the method terminates. Otherwise, $\Phi_k(\hc_k)=(\ct,q_{k+1})$, where $q_{k+1}\in\QQ$.

Note that since $F\in\ER$, Definition~\ref{def:query} gives $\QQ_F=\QQ$. Thus, the response $\eta_{k+1}=E_F(q_{k+1})$ is well defined, and we can append $(z_{k+1},r_{k+1},\eta_{k+1})$ to the record $\Rec_k$, obtaining an admissible prefix $\Rec_{k+1}$.

If no output has occurred earlier, this construction reaches an admissible prefix of length $d$. Since $\hc_d\in D_d$ and $\Phi_d$ only makes output decisions, there exists $a\in\R$ such that $\Phi_d(\hc_d)=(\st,a)$. Hence a valid execution is obtained for every $x\in\R$, and $\Omega_{T,F}=\R$.
\end{proof}

\begin{example}[Newton's method as a partial method]\label{ex:newton}
Consider the method $T$ with budget $d=2$. We define \[
D_0=\R,\qquad D_1=\R^2,\qquad
D_2=\{(x,\eta_1,\eta_2)\in\R^3:\eta_2\ne0\}
\] and the mappings \[
\begin{aligned}
\Phi_0(x)&=(\ct,(x,0)), & \Phi_1(x,\eta_1)&=(\ct,(x,1)),\\
\Phi_2(x,\eta_1,\eta_2)&=\left(\st,x-\frac{\eta_1}{\eta_2}\right).
\end{aligned}
\]

Let $f\in C^\infty(I_f)$. Note that the first query is made at the node $x$, so a valid execution requires $x\in I_f$.

Let then $x\in I_f$. By the definition of $\Phi_0$, the first query is $q_1=(x,0)$ and its response is given by \[
\eta_1=E_f(q_1)=f(x).
\] Since $(x,\eta_1)\in D_1$, we can make the second query. By the definition of $\Phi_1$, we have $q_2=(x,1)$ and \[
\eta_2=E_f(q_2)=f'(x).
\] Thus, we obtain the record \[
\Rec_2=
\bigl((x,0,f(x)),(x,1,f'(x))\bigr).
\]

For a valid execution to exist, we must have \[
(x,\eta_1,\eta_2)\in D_2,
\] which is equivalent to $f'(x)\ne0$. In this case, \[
\Phi_2(x,f(x),f'(x))
=\left(\st,x-\frac{f(x)}{f'(x)}\right).
\] Therefore, the mapping defined by the method is \begin{equation}\label{eq:newton-iteration}
T_f(x)=x-\frac{f(x)}{f'(x)}.
\end{equation} This is Newton's iteration.

We conclude that \[
\Omega_{T,f}
=\{x\in I_f:f'(x)\ne0\},
\qquad
\nu_{T,f}(x)=2
\quad\text{for every }x\in\Omega_{T,f}.
\] Note that both queries are made at the same node $x$, one to evaluate $f(x)$ and the other to evaluate $f'(x)$. Moreover, the method is partially defined, since $D_2\ne H_2$.
\end{example}

\subsection{Exact Preservation of Data}

Let $\Rec_k=((z_j,r_j,\eta_j))_{j=1}^k$ be a list. We say that a function $g$ is \emph{compatible} with $\Rec_k$ if, for every $1\le j\le k$, \[
(z_j,r_j)\in\QQ_g,\qquad
E_g(z_j,r_j)=\eta_j.
\] Likewise, we say that two functions \emph{agree on the list} $\Rec_k$ if both are compatible with it.

\begin{lemma}[Exact preservation]\label{lem:preserve}
Let $\Rec_k$ be an admissible prefix of $T$ on $(f,x)$, and suppose that $g$ is compatible with $\Rec_k$. Then $\Rec_k$ is also an admissible prefix on $(g,x)$. Moreover, if $(\Rec_k,a)$ is a valid execution on $(f,x)$, it is also valid on $(g,x)$ and \begin{equation}\label{eq:preserve-output}
T_g(x)=a=T_f(x).
\end{equation} holds.
\end{lemma}

\begin{proof}
Let $\Rec_k=((z_j,r_j,\eta_j))_{j=1}^k$ and denote by $\hc_i^f$ and $\hc_i^g$ the data vectors corresponding to each function. We prove by induction that the queries and their responses coincide up to level $k$.

Initially, we have $\hc_0^f=\hc_0^g=(x)$. Suppose that, for some $0\le i<k$, the data coincide up to level $i$, that is, $\hc_i^f=\hc_i^g$. Since $\Rec_k$ is admissible on $(f,x)$, we have \[
\hc_i^f\in D_i,\qquad
\Phi_i(\hc_i^f)=(\ct,q_{i+1}),\qquad
\eta_{i+1}=E_f(q_{i+1}).
\] Therefore, \[
\hc_i^g=\hc_i^f\in D_i,\qquad
\Phi_i(\hc_i^g)=\Phi_i(\hc_i^f)
=(\ct,q_{i+1}).
\] Moreover, since $g$ is compatible with $\Rec_k$, we have $q_{i+1}\in\QQ_g$ and \[
E_g(q_{i+1})=\eta_{i+1}=E_f(q_{i+1}).
\] Thus, the query made on $g$ is the same as on $f$, and both produce the same response. Consequently, \[
\hc_{i+1}^g
=(\hc_i^g,\eta_{i+1})
=(\hc_i^f,\eta_{i+1})
=\hc_{i+1}^f.
\] We conclude by induction that $\Rec_k$ is an admissible prefix on $(g,x)$.

Now suppose that $(\Rec_k,a)$ is a valid execution on $(f,x)$. By definition, we have \[
\hc_k^f\in D_k,\qquad
\Phi_k(\hc_k^f)=(\st,a).
\] Since $\hc_k^g=\hc_k^f$, it follows that \[
\hc_k^g\in D_k,\qquad
\Phi_k(\hc_k^g)=\Phi_k(\hc_k^f)=(\st,a).
\] Hence $(\Rec_k,a)$ is also a valid execution on $(g,x)$, and~\eqref{eq:preserve-output} holds. Note that this conclusion follows solely from the equality of the queries and their responses, without requiring continuity of the method's rules.
\end{proof}

\par\medskip
\needspace{8\baselineskip}

\section{Procedures and Valid Computations}\label{sec:procedimientos}

In a procedure, we distinguish internal, query, output, and blocking states. The general definitions apply to functions in $C^\infty(I_f)$; the auxiliary class $\DD$ will be used in the examples and local implementations.

A root $\alpha$ is called simple when $f(\alpha)=0$ and $f'(\alpha)\ne0$.

\subsection{The Local Class of Analytic Problems}

\begin{definition}[Local class $\DD$]
\label{def:localclass}
We denote by $\DD$ the class of triples $(I_f,f,\alpha_f)$ such that $I_f\subseteq\R$ is an open interval, $f:I_f\to\R$ is a real-analytic function, and \[
\alpha_f\in I_f,\qquad
f(\alpha_f)=0,\qquad
f'(u)\ne0
\quad\text{for every }u\in I_f.
\]

Each triple $(I_f,f,\alpha_f)$ constitutes a problem in the class $\DD$. When there is no risk of confusion, we write $f\in\DD$, keeping in mind that the interval $I_f$ is part of the problem.

We consider methods that use the same tuple of rules for every problem in $\DD$. Neither the interval $I_f$ nor the root $\alpha_f$ is supplied to the method as data. The queries available for the function $f$ are precisely those belonging to \[
\QQ_f=I_f\times\Nn.
\]
\end{definition}

\begin{lemma}[Uniqueness of the root in the local class]
\label{lem:localunique}
Let $(I_f,f,\alpha_f)\in\DD$. Then $\alpha_f$ is the unique root of $f$ in $I_f$, and it is simple.
\end{lemma}

\begin{proof}
By Definition~\ref{def:localclass}, $f'$ is continuous and never vanishes on the interval $I_f$. The intermediate value theorem implies that it has constant sign. Let $a,b\in I_f$ with $a<b$. By the mean value theorem, there exists $c\in(a,b)$ such that \[
f(b)-f(a)=f'(c)(b-a).
\] Since the right-hand side has a constant nonzero sign, $f$ is strictly monotone and has at most one root. Finally, $f(\alpha_f)=0$ and $f'(\alpha_f)\ne0$ show that $\alpha_f$ is that unique root and that it is simple.
\end{proof}

\subsection{Procedure and Valid Computation}

We use the sets $\QQ$, $\QQ_f$, and the mapping $E_f$ defined in Definition~\ref{def:query}.

For a finite list $b$, we denote by $b\Vert\eta$ the list obtained by appending the element $\eta$ to the end of $b$. If $b$ is a list of blocks, $b\Vert B$ denotes the list obtained by appending $B$ as a new block.

\begin{definition}[Computational procedure]
\label{def:calc}
A \emph{computational procedure} is a tuple \[
\mathscr C=
(\Sigma,\Sigma_{\rm int},
\Sigma_{\rm con},\Sigma_{\rm sal},
D_\iota,D_V,\iota,U,Q,V,A),
\] whose elements satisfy the following conditions.

The set $\Sigma$ consists of the states of the procedure, and \[
\Sigma_{\rm int},\,
\Sigma_{\rm con},\,
\Sigma_{\rm sal}\subseteq\Sigma
\] are pairwise disjoint subsets. Their elements are called internal, query, and output states, respectively.

We define \[
\Sigma_{\rm blo}:=
\Sigma\setminus
\bigl(
\Sigma_{\rm int}\cup
\Sigma_{\rm con}\cup
\Sigma_{\rm sal}
\bigr).
\] The elements of $\Sigma_{\rm blo}$ are called blocking states. Thus, \[
\Sigma=
\Sigma_{\rm int}\,\dot\cup\,
\Sigma_{\rm con}\,\dot\cup\,
\Sigma_{\rm sal}\,\dot\cup\,
\Sigma_{\rm blo},
\] where $\dot\cup$ denotes disjoint union.

The sets $D_\iota$ and $D_V$ satisfy \[
D_\iota\subseteq\R,\qquad
D_V\subseteq\Sigma_{\rm con}\times\R,
\] and the mappings of the procedure are \[
\begin{aligned}
\iota &:D_\iota\longrightarrow\Sigma, & U &: \Sigma_{\rm int}\longrightarrow\Sigma,\\
Q &: \Sigma_{\rm con}\longrightarrow\QQ, & V &:D_V\longrightarrow\Sigma,\\
A &: \Sigma_{\rm sal}\longrightarrow\R.
\end{aligned}
\]

The mapping $\iota$ determines the initial state from $x\in D_\iota$, while $U$ determines the next state from an internal state.

The mappings $Q$, $V$, and $A$ determine, respectively, the query at a query state, the next state after incorporating an admissible response, and the output.

All sets and mappings forming $\mathscr C$ are fixed independently of the function $f$. In particular, initialization $\iota(x)$ depends only on $x$; it receives no information from previous computations, nor the interval $I_f$, the root $\alpha_f$, or a description of $f$.

Continuity or computability of the mappings defining the procedure is not required.
\end{definition}

\begin{example}[A procedure for Newton's method]
\label{ex:newtoncalc}

Consider Newton's iteration given by~\eqref{eq:newton-iteration}, which we denote here by $\phi_f$, with $x\in I_f$ and $f'(x)\ne0$. Let us construct a procedure $\mathscr C$ that computes $\phi_f(x)$ using two queries.

For $x,y,z,a\in\R$, define the states \[
\begin{aligned}
c_0(x)&=(0,x),&
c_1(x,y)&=(1,x,y),\\
i(x,y,z)&=(2,x,y,z),&
o(a)&=(3,a),&
b&=(4,0).
\end{aligned}
\] The first component of each state distinguishes the different classes.

The sets and mappings of $\mathscr C$ are defined in the following table.

\begin{center}
\small
\renewcommand{\arraystretch}{1.4}
\begin{tabular}{@{}l p{0.77\linewidth}@{}}
\toprule
\textbf{Element} & \textbf{Definition}\\
\midrule
$\Sigma_{\rm con}$ & $\{c_0(x):x\in\R\}
   \cup\{c_1(x,y):x,y\in\R\}$\\

$\Sigma_{\rm int}$
& $\{i(x,y,z):x,y,z\in\R\}$\\

$\Sigma_{\rm sal}$
& $\{o(a):a\in\R\}$\\

$\Sigma_{\rm blo}$
& $\{b\}$\\

$\Sigma$
& $\Sigma_{\rm con}\,\dot\cup\,
   \Sigma_{\rm int}\,\dot\cup\,
   \Sigma_{\rm sal}\,\dot\cup\,
   \Sigma_{\rm blo}$\\

$D_\iota$
& $\R$\\

$D_V$
& $\Sigma_{\rm con}\times\R$\\

$\iota$
& $\iota(x)=c_0(x)$\\

$Q$
& $\begin{aligned}
Q(c_0(x))&=(x,0),\\
Q(c_1(x,y))&=(x,1)
\end{aligned}$\\

$V$
& $\begin{aligned}
V(c_0(x),y)&=c_1(x,y),\\
V(c_1(x,y),z)&=i(x,y,z)
\end{aligned}$\\

$U$
& $\displaystyle
U(i(x,y,z))=
\begin{cases}
o(x-y/z),&z\ne0,\\
b,&z=0
\end{cases}$\\

$A$
& $A(o(a))=a$\\
\bottomrule
\end{tabular}
\end{center}

Note that the four classes of states are disjoint and that all the mappings have the domains and codomains required by Definition~\ref{def:calc}. Therefore, $\mathscr C$ is a computational procedure.

Let us see how it acts on a function $f\in C^\infty(I_f)$. Fix $x\in I_f$ with $f'(x)\ne0$. By the preceding definitions, \[
\begin{aligned}
s_0&=\iota(x)=c_0(x), & Q(s_0)&=(x,0),\\
E_f(Q(s_0))&=f(x).
\end{aligned}
\] Incorporating this response gives \[
s_1=V(s_0,f(x))=c_1(x,f(x)).
\]

In the new state, we have \[
Q(s_1)=(x,1),\qquad
E_f(Q(s_1))=f'(x),
\] and therefore \[
s_2=V(s_1,f'(x))
=i(x,f(x),f'(x)).
\]

Since $f'(x)\ne0$, the internal transformation produces \[
s_3=U(s_2)
=o\left(x-\frac{f(x)}{f'(x)}\right).
\] Thus, $s_3\in\Sigma_{\rm sal}$ and, by~\eqref{eq:newton-iteration}, the output satisfies $A(s_3)=\phi_f(x)$.

The queries in the execution are \begin{equation}\label{eq:newton-query-pairs}
q_1=(x,0),\qquad q_2=(x,1).
\end{equation} Their responses are $f(x)$ and $f'(x)$; division is performed in the internal transition $U$, not through another query.

If $x\in I_f$ and $f'(x)=0$, both queries remain available, but \[
U(i(x,f(x),0))=b
\in\Sigma_{\rm blo},
\] so no output is obtained. Likewise, if $x\notin I_f$, then \[
Q(\iota(x))=(x,0)\notin\QQ_f,
\] and the first query has no response.

In particular, for every problem $(I_f,f,\alpha_f)\in\DD$ and $x\in I_f$, we have $f'(x)\ne0$. Therefore, the procedure reaches an output state and computes $\phi_f(x)$.

\medskip

Let us now compare this construction with the method having partial rules in Example~\ref{ex:newton}. In the previous model, the rules $\Phi_0$, $\Phi_1$, and $\Phi_2$ of Example~\ref{ex:newton} were used; they determine exactly the queries shown in~\eqref{eq:newton-query-pairs}.

Both models obtain the same responses, $y=f(x)$ and $z=f'(x)$, and the same output, given by~\eqref{eq:newton-iteration}.

The difference lies in the description of the computation. In the previous model, the rules $\Phi_k$ directly determine the queries or the output. In the new procedure, these operations are represented by the mappings $Q$, $V$, $U$, and $A$.

In particular, the division appearing in Newton's formula is performed by $U$, without introducing a new query. Therefore, the procedure uses two evaluations of the function and its derivatives, regardless of the internal operations needed to obtain the output.
\end{example}

\begin{definition}[Valid computation]
\label{def:validcalc}
Let $\mathscr C$ be a computational procedure, let $f\in C^\infty(I_f)$ and $x\in\R$. A finite sequence of states \[
\boldsymbol s=(s_0,\ldots,s_L),
\qquad L\in\Nn,\quad s_i\in\Sigma,
\] is called a \emph{valid computation} of $\mathscr C$ on $(f,x)$ if \begin{equation}\label{eq:calcstart}
x\in D_\iota,\qquad
s_0=\iota(x),\qquad
s_L\in\Sigma_{\rm sal},
\end{equation} and, for each $0\le i<L$, one of the following conditions holds: \begin{equation}\label{eq:calctransitions}
\begin{cases}
s_i\in\Sigma_{\rm int},\\
s_{i+1}=U(s_i),
\end{cases}
\quad\text{or}\quad
\begin{cases}
s_i\in\Sigma_{\rm con},\\
Q(s_i)\in\QQ_f,\\
(s_i,E_f(Q(s_i)))\in D_V,\\
s_{i+1}=V(s_i,E_f(Q(s_i))).
\end{cases}
\end{equation} These alternatives are mutually exclusive, since $\Sigma_{\rm int}$ and $\Sigma_{\rm con}$ are disjoint.

The \emph{output} of the computation is $A(s_L)$. We say that $\mathscr C$ \emph{terminates with output} $a\in\R$ on $(f,x)$ if there exists a valid computation $\boldsymbol s$ such that \[
A(s_L)=a.
\]

We say that the computation of $\mathscr C$ on $(f,x)$ is \emph{undefined} if no finite sequence satisfies \eqref{eq:calcstart} and \eqref{eq:calctransitions}.

In particular, a query $Q(s_i)\notin\QQ_f$ provides no response and does not allow the computation to continue. Nor is it regarded as a failure signal that the procedure could use.
\end{definition}

By~\eqref{eq:calcstart} and~\eqref{eq:calctransitions}, all states preceding the last are internal or query states, and the last is an output state. Since these three classes are disjoint from $\Sigma_{\rm blo}$, no blocking state appears in a valid computation.

\begin{lemma}[Uniqueness of the computation]
\label{lem:calcunique}
Let $\mathscr C$ be a computational procedure, let $f\in C^\infty(I_f)$ and $x\in\R$. Then there exists at most one valid computation of $\mathscr C$ on $(f,x)$.
\end{lemma}

\begin{proof}
Let \[
\boldsymbol s=(s_0,\ldots,s_L),
\qquad
\boldsymbol t=(t_0,\ldots,t_M)
\] be two valid computations of $\mathscr C$ on $(f,x)$.

By \eqref{eq:calcstart}, \[
s_0=\iota(x)=t_0.
\] We prove by induction that \begin{equation}\label{eq:calc-unique-induction}
s_i=t_i,\qquad 0\le i\le\min\{L,M\}.
\end{equation}

Suppose that $s_i=t_i$ for some $i<\min\{L,M\}$. By \eqref{eq:calctransitions}, \[
s_i=t_i\in
\Sigma_{\rm int}\cup\Sigma_{\rm con}.
\]

If $s_i\in\Sigma_{\rm int}$, then \[
s_{i+1}=U(s_i)=U(t_i)=t_{i+1}.
\]

If $s_i\in\Sigma_{\rm con}$, we have \[
q:=Q(s_i)=Q(t_i)\in\QQ_f,
\qquad y:=E_f(q),
\] and \[
(s_i,y)=(t_i,y)\in D_V.
\] Therefore, \[
s_{i+1}
=V(s_i,y)
=V(t_i,y)
=t_{i+1}.
\] Thus, in both cases, $s_{i+1}=t_{i+1}$, and the induction establishes~\eqref{eq:calc-unique-induction}.

We show that $L=M$. If $L<M$, then, by \eqref{eq:calcstart} and \eqref{eq:calctransitions}, \[
s_L\in\Sigma_{\rm sal},
\qquad
t_L\in\Sigma_{\rm int}\cup\Sigma_{\rm con}.
\] But $s_L=t_L$, which contradicts \[
\Sigma_{\rm sal}\cap
(\Sigma_{\rm int}\cup\Sigma_{\rm con})
=\varnothing.
\] The case $M<L$ is ruled out in the same way. Therefore, $L=M$ and, by~\eqref{eq:calc-unique-induction}, the two computations coincide, including their output $A(s_L)=A(t_L)$.

\end{proof}

By Lemma~\ref{lem:calcunique}, there exists at most one valid computation for each $(f,x)$. Therefore, we may define \[
\Omega_{\mathscr C,f}
:=
\left\{
x\in\R:
\text{there exists a valid computation of }
\mathscr C\text{ on }(f,x)
\right\},
\] and the mapping \[
C_f:\Omega_{\mathscr C,f}\longrightarrow\R,
\qquad
C_f(x):=A(s_L),
\] where $(s_0,\ldots,s_L)$ is the unique valid computation of $\mathscr C$ on $(f,x)$.

\subsection{Evaluations Used and Realized Cost}

\begin{definition}[Evaluation used and computation record]
\label{def:used}
Let $\mathscr C$ be a computational procedure, let $f\in C^\infty(I_f)$ and $x\in\Omega_{\mathscr C,f}$. Consider the unique valid computation $\boldsymbol s=(s_0,\ldots,s_L)$.

\begin{enumerate}
\item[(a)] \textbf{Evaluations used.}
We define \[
\mathcal J(\boldsymbol s):=
\{i\in\{0,\ldots,L-1\}:
s_i\in\Sigma_{\rm con}\}.
\] For each $i\in\mathcal J(\boldsymbol s)$, we have \[
q_i:=Q(s_i)=(z_i,r_i)\in\QQ_f,
\qquad
\eta_i:=E_f(q_i)=f^{(r_i)}(z_i).
\] The triple $(i,q_i,\eta_i)$ is called an \emph{evaluation used}. It is counted regardless of whether its response enters the final output.

\item[(b)] \textbf{Evaluation record.}
If \begin{equation}\label{eq:query-index-list}
\mathcal J(\boldsymbol s)=\{i_1<\cdots<i_k\}.
\end{equation} We define \[
\Rec_{\mathscr C,f}(x):=
\bigl(
(z_{i_j},r_{i_j},\eta_{i_j})
\bigr)_{j=1}^k.
\] The record preserves the order in which evaluations are performed. If $k=0$, the record is the empty list.

\item[(c)] \textbf{Number of evaluations.}
We define \[
N_{\mathscr C}(f,x):=k
\] and, for each $r\in\Nn$, \[
N_{\mathscr C,r}(f,x):=
\bigl|\{i\in\mathcal J(\boldsymbol s):
r_i=r\}\bigr|.
\]

\item[(d)] \textbf{Repetition and reuse.}
Two evaluations performed at distinct indices are counted separately, even if they correspond to the same query: \[
i\ne j,\quad q_i=q_j
\quad\Longrightarrow\quad
\text{two evaluations used}.
\] By contrast, internal transformations do not add evaluations, since \[
s_i\in\Sigma_{\rm int}
\quad\Longrightarrow\quad
i\notin\mathcal J(\boldsymbol s).
\]
\end{enumerate}
\end{definition}

\begin{lemma}[Partition by orders]
\label{lem:countpartition}
Let $\mathscr C$ be a computational procedure, let $f\in C^\infty(I_f)$ and $x\in\Omega_{\mathscr C,f}$. Then \begin{equation}\label{eq:countpartition}
N_{\mathscr C}(f,x)
=\sum_{r\in\Nn}N_{\mathscr C,r}(f,x).
\end{equation}

Likewise, let $T$ be a method and $x\in\Omega_{T,f}$. Consider the record of its valid execution, which we write as \begin{equation}\label{eq:generic-T-register}
\Rec_{T,f}(x)=\bigl((z_j,r_j,\eta_j)\bigr)_{j=1}^k.
\end{equation} We define \[
\nu_{T,f}(x):=k,
\qquad
\nu^{(r)}_{T,f}(x):=
\bigl|\{j\in\{1,\ldots,k\}:r_j=r\}\bigr|,
\quad r\in\Nn.
\] Then \[
\nu_{T,f}(x)
=\sum_{r\in\Nn}\nu^{(r)}_{T,f}(x).
\]
\end{lemma}

\begin{proof}
Let $\boldsymbol s=(s_0,\ldots,s_L)$ be the valid computation of $\mathscr C$ on $(f,x)$, and use the indices $i_1<\cdots<i_k$ from~\eqref{eq:query-index-list}. For each $r\in\Nn$, define \[
J_r:=\{j\in\{1,\ldots,k\}:r_{i_j}=r\}.
\]

Since each query has a unique derivative order, we have \[
J_r\cap J_s=\varnothing
\quad(r\ne s),
\qquad
\bigcup_{r\in\Nn}J_r=\{1,\ldots,k\}.
\] Moreover, $|\{r\in\Nn:J_r\ne\varnothing\}|\le k,$ so the preceding union contains only finitely many nonempty sets. Thus, \[
\begin{aligned}
N_{\mathscr C}(f,x)
&=k\\
&=\sum_{r\in\Nn}|J_r|\\
&=\sum_{r\in\Nn}N_{\mathscr C,r}(f,x),
\end{aligned}
\] which proves \eqref{eq:countpartition}.

For the method $T$, let $J_r^T:=\{j\in\{1,\ldots,k\}:r_j=r\}$, with the record in~\eqref{eq:generic-T-register}. These sets also form a finite partition of $\{1,\ldots,k\}$. Since $\nu^{(r)}_{T,f}(x)=|J_r^T|$, the same counting argument proves the second identity in the lemma.

\end{proof}

\subsection{Local Implementations, Budgets, and Profiles}

The following notions describe the cost and profile of a fixed local implementation, without minimizing over different procedures.

\begin{definition}[Local implementation, total cost, and profile]
\label{def:realizedcost}\label{def:operative}

\begin{enumerate}

\item[(a)] \textbf{Mapping and domain.}

Let $\phi$ be an assignment that associates with each problem $f\in\DD$ a partial mapping \[
\phi(f):D_{\phi,f}\longrightarrow\R,
\qquad D_{\phi,f}\subseteq\R,
\] whose domain contains a neighborhood of $\alpha_f$. We denote the domain of that mapping by \[
\Dom\phi(f):=D_{\phi,f}
\]

\item[(b)] \textbf{Local input set.}

A \emph{local input set} is a set $\mathcal U$ of pairs $(f,x)$ satisfying \begin{equation}\label{eq:localset}
\begin{gathered}
\mathcal U\subseteq
\bigl\{
(f,x):f\in\DD,\,
x\in\Dom\phi(f),\,
x\ne\alpha_f
\bigr\},\\
\forall f\in\DD\
\exists\delta_f>0\
\forall x\in\R,\\
0<|x-\alpha_f|<\delta_f
\ \Longrightarrow\
(f,x)\in\mathcal U.
\end{gathered}
\end{equation}

\item[(c)] \textbf{Local implementation.}

A \emph{local implementation of $\phi$} is a triple \[
\mathfrak C=(\phi,\mathscr C,\mathcal U),
\] where $\mathscr C$ is a computational procedure fixed for all problems in $\DD$, and $\mathcal U$ is a local input set such that \begin{equation}\label{eq:realization}
\forall(f,x)\in\mathcal U,\qquad
x\in\Omega_{\mathscr C,f},
\quad
C_f(x)=\phi(f)(x).
\end{equation}

\item[(d)] \textbf{Total cost and budgets.}

The \emph{total cost} of the implementation $\mathfrak C$ is defined by \begin{equation}\label{eq:realizedcost}
v(\mathfrak C):=
\sup_{(f,x)\in\mathcal U}
N_{\mathscr C}(f,x)
\in[0,+\infty].
\end{equation} We also write \[
v(\phi;\mathscr C,\mathcal U)
:=v(\mathfrak C).
\]

An integer $n\in\Nn$ is a \emph{local budget} for $\mathfrak C$ if \[
v(\mathfrak C)\le n.
\]

We say that $n$ is a \emph{global budget} of $\mathscr C$ on the class $\DD$ if \[
\forall f\in\DD\
\forall x\in\Omega_{\mathscr C,f},
\qquad N_{\mathscr C}(f,x)\le n.
\]

\item[(e)] \textbf{Cost by orders and fixed profile.}

For each $r\in\Nn$, we define \[
b_r(\mathfrak C):=
\sup_{(f,x)\in\mathcal U}
N_{\mathscr C,r}(f,x).
\]

We say that $\mathfrak C$ has \emph{fixed profile} $(c_r)_{r\in\Nn}$, with $c_r\in\Nn$, if \[
\forall(f,x)\in\mathcal U\
\forall r\in\Nn,\qquad
N_{\mathscr C,r}(f,x)=c_r.
\] In this case, we write \[
v_r(\mathfrak C):=c_r.
\]

If the procedure admits different computation branches, the total cost $v(\mathfrak C)$ is not defined by the sum \[
\sum_{r\in\Nn}b_r(\mathfrak C).
\]
\end{enumerate}
\end{definition}

\begin{remark}[Dependence of the cost]
Note that the cost $v(\mathfrak C)$ depends on the procedure $\mathscr C$ and the local input set $\mathcal U$. In particular, two procedures computing the same mapping $\phi$ can have different costs.

On the other hand, $\mathcal U$ is not a datum supplied to the procedure, and the radii $\delta_f$ in \eqref{eq:localset} only guarantee its local character. In particular, none of these objects is used as an argument of the mappings \[
\iota,\quad U,\quad Q,\quad V,\quad A.
\]

Thus, the cost $v(\mathfrak C)$ is not a minimum over implementations of $\phi$.
\end{remark}

\begin{example}[Local implementations of Newton's method]
\label{ex:localrealization}

For each $(I_f,f,\alpha_f)\in\DD$, consider Newton's mapping $\phi(f):I_f\to\R$, whose expression is given in~\eqref{eq:newton-iteration}. Note that $\phi(f)$ is well defined, since \[
f'(x)\ne0,\qquad x\in I_f.
\]

We define the local input set \begin{equation}\label{eq:newton-local-U}
\mathcal U:=\{(f,x):f\in\DD,\,x\in I_f\setminus\{\alpha_f\}\}.
\end{equation} Since $I_f$ is open and $\alpha_f\in I_f$, for each $f\in\DD$ there exists $\delta_f>0$ such that \[
(\alpha_f-\delta_f,
\alpha_f+\delta_f)\subseteq I_f.
\] Therefore, \[
0<|x-\alpha_f|<\delta_f
\quad\Longrightarrow\quad
(f,x)\in\mathcal U,
\] and $\mathcal U$ is a local input set.

\medskip
\noindent\textbf{(a) Implementation with a fixed profile.}

Let $\mathscr C_1$ be Newton's procedure from Example~\ref{ex:newtoncalc}. For every $(f,x)\in\mathcal U$, its queries and responses are those of Example~\ref{ex:newtoncalc}, given in~\eqref{eq:newton-query-pairs}. Since $f'(x)\ne0$, the computation is valid and, by~\eqref{eq:newton-iteration}, $C_f^{(1)}(x)=\phi(f)(x)$. Thus, $\mathfrak C_1:=(\phi,\mathscr C_1,\mathcal U)$ is a local implementation of $\phi$.

For each input in $\mathcal U$, the record contains one query of order zero and another of order one. Thus, $\mathfrak C_1$ has fixed profile \[
(c_r)_{r\ge0}=(1,1,0,0,\ldots),\qquad v(\mathfrak C_1)=2.
\] In particular, $b_0(\mathfrak C_1)=b_1(\mathfrak C_1)=1$ and $b_r(\mathfrak C_1)=0$ for $r\ge2$.

In particular, $n=2$ is a local budget for $\mathfrak C_1$ and also a global budget for $\mathscr C_1$ on $\DD$.

\medskip
\noindent\textbf{(b) An implementation with different branches.}

Let us construct another procedure $\mathscr C_2$ computing the same mapping $\phi$, but making an additional query according to the sign of $f(x)$.

We define \[
\rho(y):=
\begin{cases}
2,&y>0,\\
3,&y\le0.
\end{cases}
\] The procedure successively makes the queries \[
q_1=(x,0),\qquad q_2=(x,\rho(f(x))),\qquad q_3=(x,1),
\] and produces Newton's output defined in~\eqref{eq:newton-iteration}.

To specify this construction, consider the labeled states \[
\begin{aligned}
c_0(x)&=(0,x), & c_1(x,y)&=(1,x,y),\\
c_2(x,y,w)&=(2,x,y,w), & i(x,y,w,z)&=(3,x,y,w,z),\\
o(a)&=(4,a), & b&=(5,0).
\end{aligned}
\] We define \[
\begin{aligned}
\Sigma_{\rm con}
&=\{c_0(x),c_1(x,y),c_2(x,y,w):
x,y,w\in\R\},\\
\Sigma_{\rm int}
&=\{i(x,y,w,z):x,y,w,z\in\R\},\\
\Sigma_{\rm sal}
&=\{o(a):a\in\R\},\\
\Sigma_{\rm blo}&=\{b\},
\end{aligned}
\] and let $\Sigma$ be their disjoint union.

We take \[
D_\iota=\R,\qquad
D_V=\Sigma_{\rm con}\times\R,
\qquad
\iota(x)=c_0(x),
\] with \[
\begin{aligned}
Q(c_0(x))&=(x,0), & Q(c_1(x,y))&=(x,\rho(y)),\\
Q(c_2(x,y,w))&=(x,1).
\end{aligned}
\] \[
\begin{aligned}
V(c_0(x),y)&=c_1(x,y), & V(c_1(x,y),w)&=c_2(x,y,w),\\
V(c_2(x,y,w),z)&=i(x,y,w,z).
\end{aligned}
\] and \[
U(i(x,y,w,z))=
\begin{cases}
o(x-y/z),&z\ne0,\\
b,&z=0,
\end{cases}
\qquad
A(o(a))=a.
\] These mappings define a procedure $\mathscr C_2$ in the sense of Definition~\ref{def:calc}.

For $(f,x)\in\mathcal U$, we have \[
f(x)\ne0,\qquad f'(x)\ne0.
\] Moreover, all three queries belong to $\QQ_f$, and the computation terminates with the same output $C_f^{(2)}(x)=\phi(f)(x)$, by~\eqref{eq:newton-iteration}. Therefore, $\mathfrak C_2:=(\phi,\mathscr C_2,\mathcal U)$ is another local implementation of $\phi$.

In this case, \[
N_{\mathscr C_2}(f,x)=3,
\qquad (f,x)\in\mathcal U,
\] but the evaluation counts by order depend on the branch: \[
\begin{array}{c|cccc|c}
&N_0&N_1&N_2&N_3&N\\ \hline
f(x)>0&1&1&1&0&3\\
f(x)<0&1&1&0&1&3
\end{array}
\] where $N_r=N_{\mathscr C_2,r}(f,x)$ and $N=N_{\mathscr C_2}(f,x)$.

Since $f$ is strictly monotone and $f(\alpha_f)=0$, there are points in $I_f$ on both sides of $\alpha_f$ at which $f$ takes opposite signs. Therefore, both branches occur in $\mathcal U$, and \[
b_0(\mathfrak C_2)
=b_1(\mathfrak C_2)
=b_2(\mathfrak C_2)
=b_3(\mathfrak C_2)=1,
\] \[
b_r(\mathfrak C_2)=0,\qquad r\ge4.
\]

Consequently, \[
v(\mathfrak C_2)=3
<4=\sum_{r\ge0}b_r(\mathfrak C_2).
\] The table shows that $\mathfrak C_2$ does not have a fixed profile: its number of order-two evaluations depends on the sign of $f(x)$. The implementations $\mathfrak C_1$ and $\mathfrak C_2$ compute the same mapping on $\mathcal U$, but have costs $2$ and $3$, respectively.
\end{example}

\begin{lemma}[Finite cost, bounds by order, and fixed profile]
\label{lem:costprofile}
Let $\mathfrak C=(\phi,\mathscr C,\mathcal U)$ be a local implementation. Then the following properties hold:

\begin{enumerate}
\item[(a)] If $v(\mathfrak C)<+\infty$, then $v(\mathfrak C)\in\Nn$ and there exists $(f_0,x_0)\in\mathcal U$ such that \[
v(\mathfrak C)=N_{\mathscr C}(f_0,x_0).
\]

\item[(b)] For every $r\in\Nn$, \begin{equation}\label{eq:summax}
b_r(\mathfrak C)\le v(\mathfrak C)
\le\sum_{s\in\Nn}b_s(\mathfrak C),
\end{equation} where the sum may be infinite.

\item[(c)] If $\mathfrak C$ has fixed profile $(c_r)_{r\in\Nn}$, then \[
\bigl|\{r\in\Nn:c_r\ne0\}\bigr|<\infty
\] and \begin{equation}\label{eq:fixedprofile}
v(\mathfrak C)
=\sum_{r\in\Nn}v_r(\mathfrak C)
=\sum_{r\in\Nn}c_r.
\end{equation}
\end{enumerate}
\end{lemma}

\begin{proof}
\noindent\textbf{(a) Finite cost.}

Note that $\mathcal U\ne\varnothing$. Indeed, $\bigl((-1,1),u\mapsto u,0\bigr)\in\DD,$ and, by \eqref{eq:localset}, there exist $f\in\DD$ and $x\in\R$ such that $(f,x)\in\mathcal U$.

Suppose that $M:=v(\mathfrak C)<+\infty$ and define \[
S:=\{N_{\mathscr C}(f,x):
(f,x)\in\mathcal U\}.
\] Since $N_{\mathscr C}(f,x)\in\Nn$, we have $\varnothing\ne S \subseteq\{0,1,\ldots,\lfloor M\rfloor\}.$ Therefore, $S$ is a finite set and $v(\mathfrak C)=\sup S=\max S\in\Nn.$ Consequently, there exists $(f_0,x_0)\in\mathcal U$ such that \[
v(\mathfrak C)
=N_{\mathscr C}(f_0,x_0),
\] which proves (a).

\medskip
\noindent\textbf{(b) Bounds by order.}

By Lemma~\ref{lem:countpartition}, for every $(f,x)\in\mathcal U$ and $r\in\Nn$, \[
\begin{aligned}
N_{\mathscr C,r}(f,x)
&\le N_{\mathscr C}(f,x)\\
&=\sum_{s\in\Nn}N_{\mathscr C,s}(f,x)\\
&\le\sum_{s\in\Nn}b_s(\mathfrak C).
\end{aligned}
\] Taking suprema over $(f,x)\in\mathcal U$, we obtain the bound~\eqref{eq:summax}. The sum on the right may be infinite, which proves (b).

\medskip
\noindent\textbf{(c) Fixed profile.}

Suppose that $\mathfrak C$ has fixed profile $(c_r)_{r\in\Nn}$. Fix $(f_0,x_0)\in\mathcal U$. Then \[
c_r=N_{\mathscr C,r}(f_0,x_0),
\qquad r\in\Nn.
\] Since the computation on $(f_0,x_0)$ uses finitely many evaluations, \[
\bigl|\{r\in\Nn:c_r\ne0\}\bigr|
\le N_{\mathscr C}(f_0,x_0)<\infty.
\] Therefore, the profile has finite support.

Moreover, for every $(f,x)\in\mathcal U$, by Lemma~\ref{lem:countpartition}, \[
\begin{aligned}
N_{\mathscr C}(f,x)
&=\sum_{r\in\Nn}N_{\mathscr C,r}(f,x)\\
&=\sum_{r\in\Nn}c_r.
\end{aligned}
\] Consequently, the supremum defining $v(\mathfrak C)$ equals this constant sum, and, since $v_r(\mathfrak C)=c_r$, we obtain~\eqref{eq:fixedprofile}. This proves (c).
\end{proof}

\par\medskip
\needspace{8\baselineskip}

\section{Exact Representation of Procedures by Queries}\label{sec:representacion}

We shall represent computations that terminate within a fixed budget, without assuming termination or a uniform cost for the original procedure.

\begin{theorem}[Exact representation of a computation with a budget]
\label{thm:exactrepresentation}
\label{lem:serialization}
\label{prop:automatic}

Let $\mathscr C$ be a computational procedure and $n\in\N$. Then there exists a method $T_{\mathscr C,n}$, in the sense of Definition~\ref{def:method}, with budget $n$, whose rules depend only on $\mathscr C$ and $n$, and which satisfies the following properties for every $f\in C^\infty(I_f)$.

\begin{enumerate}
\item[(a)] \textbf{Domain.}
\begin{equation}\label{eq:exactdomain}
\Omega_{T_{\mathscr C,n},f}
=
\left\{
x\in\Omega_{\mathscr C,f}:
N_{\mathscr C}(f,x)\le n
\right\}.
\end{equation}

\item[(b)] \textbf{Output and record.}

For every $x\in\Omega_{T_{\mathscr C,n},f}$, \begin{equation}\label{eq:exactrecords}
(T_{\mathscr C,n})_f(x)=C_f(x),
\qquad
\Rec_{T_{\mathscr C,n},f}(x)
=\Rec_{\mathscr C,f}(x).
\end{equation}

\item[(c)] \textbf{Number of evaluations.}

For every $x\in\Omega_{T_{\mathscr C,n},f}$, \begin{equation}\label{eq:exactcounts}
\begin{aligned}
\nu_{T_{\mathscr C,n},f}(x)
&=N_{\mathscr C}(f,x),\\
\nu^{(r)}_{T_{\mathscr C,n},f}(x)
&=N_{\mathscr C,r}(f,x),
\qquad r\in\Nn.
\end{aligned}
\end{equation}
\end{enumerate}

In particular, if $\mathfrak C=(\phi,\mathscr C,\mathcal U)$ is a local implementation with $v(\mathfrak C)\le n$, then \[
\forall(f,x)\in\mathcal U,\qquad
x\in\Omega_{T_{\mathscr C,n},f},
\] and \eqref{eq:exactrecords} and \eqref{eq:exactcounts} hold.
\end{theorem}

\begin{proof}
\proofstep{1. Construction of the method.}

We begin by defining a partial mapping that allows all consecutive internal states of a procedure to be combined into a single step.

For $s,t\in\Sigma$, we write $(s,t)\in\Gamma$ if there exist $m\in\Nn$ and states $s^{(0)},\ldots,s^{(m)}\in\Sigma$ such that \[
\begin{cases}
s^{(0)}=s,\quad
s^{(m)}=t\in\Sigma_{\rm con}\cup\Sigma_{\rm sal},\\
s^{(i)}\in\Sigma_{\rm int},\\
s^{(i+1)}=U(s^{(i)}),
\qquad 0\le i<m.
\end{cases}
\]

Note that, for each $s\in\Sigma$, there is at most one $t$ such that $(s,t)\in\Gamma$. Indeed, since $U$ is a mapping, two lists starting at $s$ coincide at all their common indices.

If their lengths were $m<m'$, the state at index $m$ would belong simultaneously to $\Sigma_{\rm con}\cup\Sigma_{\rm sal}$ and $\Sigma_{\rm int}$, contradicting \[
\Sigma_{\rm int}\cap
(\Sigma_{\rm con}\cup\Sigma_{\rm sal})
=\varnothing.
\] Therefore, both lists have the same length and the same final state.

Thus, the partial mapping \[
B:D_B\longrightarrow
\Sigma_{\rm con}\cup\Sigma_{\rm sal},
\] is well defined, where \[
D_B:=\{s\in\Sigma:
\exists t\in\Sigma,\ (s,t)\in\Gamma\},
\qquad
B(s):=t.
\] In particular, \[
B(s)=s,
\qquad
s\in\Sigma_{\rm con}\cup\Sigma_{\rm sal},
\] since in this case we may take $m=0$.

We now construct the rules of the method from the formal histories \[
H_j:=\R^{j+1},
\qquad 0\le j\le n.
\]

We define \[
G_0:=
\{(x)\in H_0:
x\in D_\iota,\ \iota(x)\in D_B\},
\] and \[
b_0((x)):=B(\iota(x)),
\qquad (x)\in G_0.
\]

For $1\le j\le n$, define recursively \[
\begin{aligned}
G_j:=\bigl\{
(\hc,\eta)\in H_j:\;&
\hc\in G_{j-1},\\
&b_{j-1}(\hc)\in\Sigma_{\rm con},\\
&(b_{j-1}(\hc),\eta)\in D_V,\\
&V(b_{j-1}(\hc),\eta)\in D_B
\bigr\},
\end{aligned}
\] and, for $(\hc,\eta)\in G_j$, \[
b_j((\hc,\eta))
:=B\bigl(V(b_{j-1}(\hc),\eta)\bigr).
\]

By induction, the mappings $b_j$ are well defined and \[
b_j:G_j\longrightarrow
\Sigma_{\rm con}\cup\Sigma_{\rm sal},
\qquad 0\le j\le n.
\]

For $0\le j<n$, take $D_j:=G_j$ and define \[
\Phi_j(\hc_j)=
\begin{cases}
(\ct,Q(b_j(\hc_j))),
&b_j(\hc_j)\in\Sigma_{\rm con},\\[1mm]
(\st,A(b_j(\hc_j))),
&b_j(\hc_j)\in\Sigma_{\rm sal}.
\end{cases}
\]

At the last level, define \[
D_n:=
\{\hc_n\in G_n:
b_n(\hc_n)\in\Sigma_{\rm sal}\},
\] and \[
\Phi_n(\hc_n)
=(\st,A(b_n(\hc_n))),
\qquad \hc_n\in D_n.
\]

By construction, \[
\Phi_j:D_j\longrightarrow\mathcal A
\quad(0\le j<n),
\qquad
\Phi_n:D_n\longrightarrow\mathcal A_s.
\] Therefore, these rules define a method $T_{\mathscr C,n}$ with budget $n$.

Moreover, all the preceding sets and mappings depend only on $\mathscr C$ and $n$. In particular, the rules do not depend on $f$, $I_f$, $\alpha_f$, or $\mathcal U$.

\proofstep{2. Proof of (a): equality of domains.}

We first prove that \[
\{x\in\Omega_{\mathscr C,f}:
N_{\mathscr C}(f,x)\le n\}
\subseteq\Omega_{T_{\mathscr C,n},f}.
\]

Fix $f\in C^\infty(I_f)$ and $x\in\Omega_{\mathscr C,f}$ with $N_{\mathscr C}(f,x)=k\le n$. Let \[
\boldsymbol s=(s_0,\ldots,s_L)
\] be the unique valid computation of $\mathscr C$ on $(f,x),$, and write $\mathcal J(\boldsymbol s)=\{i_1<\cdots<i_k\}$ as in~\eqref{eq:query-index-list}, with the additional convention $i_{k+1}:=L$.

If $k=0$, all states preceding $s_L$ are internal. Therefore, \[
b_0((x))=B(\iota(x))
=B(s_0)=s_L\in\Sigma_{\rm sal}.
\] Thus, the method terminates without making queries and produces the output $A(s_L)$.

Now suppose that $k\ge1$. For $1\le j\le k$, define \[
q_j:=Q(s_{i_j})=(z_j,r_j),
\qquad
\eta_j:=E_f(q_j),
\] and consider the histories \[
\hc_0:=(x),\qquad
\hc_j:=(x,\eta_1,\ldots,\eta_j).
\]

Since the states preceding $s_{i_1}$ are internal, \[
\hc_0\in G_0,\qquad
b_0(\hc_0)=s_{i_1}.
\]

We show by induction that \[
\hc_j\in G_j,\qquad
b_j(\hc_j)=s_{i_{j+1}},
\qquad 1\le j\le k.
\]

Suppose that $b_{j-1}(\hc_{j-1})=s_{i_j}$. Since $\boldsymbol s$ is a valid computation, we have \[
s_{i_j}\in\Sigma_{\rm con},
\qquad
(s_{i_j},\eta_j)\in D_V,
\] and \[
V(s_{i_j},\eta_j)=s_{i_j+1}.
\]

Moreover, only internal states occur from $s_{i_j+1}$ to $s_{i_{j+1}}$. Therefore, \[
s_{i_j+1}\in D_B,
\qquad
B(s_{i_j+1})=s_{i_{j+1}}.
\] From the definition of $G_j$ and $b_j$, we conclude that \[
\hc_j\in G_j,
\] \[
\begin{aligned}
b_j(\hc_j)
&=B\bigl(V(s_{i_j},\eta_j)\bigr)\\
&=B(s_{i_j+1})
=s_{i_{j+1}}.
\end{aligned}
\] This completes the induction.

Consequently, \[
\Phi_{j-1}(\hc_{j-1})
=(\ct,q_j),
\qquad 1\le j\le k,
\] and, since $i_{k+1}=L$, \[
b_k(\hc_k)=s_L\in\Sigma_{\rm sal}.
\]

If $k<n$, we have $\hc_k\in D_k=G_k$. If $k=n$, the last membership guarantees $\hc_n\in D_n$. In both cases, \[
\Phi_k(\hc_k)=(\st,A(s_L)).
\]

Therefore, the method makes the same $k$ queries and terminates with output $A(s_L)$. Thus, \[
x\in\Omega_{T_{\mathscr C,n},f}.
\]

We now prove the reverse inclusion. Let \[
x\in\Omega_{T_{\mathscr C,n},f},
\] and consider its valid execution \[
\Rec_k=
\bigl((z_j,r_j,\eta_j)\bigr)_{j=1}^k,
\qquad k\le n,
\] with output $a$.

For $0\le j\le k$, write \[
\hc_j=(x,\eta_1,\ldots,\eta_j),
\qquad
\sigma_j:=b_j(\hc_j).
\]

By the definition of the rules and the validity of the execution, \[
\hc_j\in G_j,\qquad 0\le j\le k,
\] \[
\sigma_{j-1}\in\Sigma_{\rm con},
\qquad
Q(\sigma_{j-1})=(z_j,r_j)\in\QQ_f,
\] \[
\eta_j=E_f(z_j,r_j),
\qquad 1\le j\le k,
\qquad\text{and}\qquad
\sigma_k\in\Sigma_{\rm sal},
\qquad A(\sigma_k)=a.
\]

Moreover, by the construction of $G_j$, \[
(\sigma_{j-1},\eta_j)\in D_V,
\qquad 1\le j\le k,
\] and, by the definition of $B$, \[
(\iota(x),\sigma_0)\in\Gamma,
\] \[
\bigl(
V(\sigma_{j-1},\eta_j),\sigma_j
\bigr)\in\Gamma,
\qquad 1\le j\le k.
\]

Each of these pairs determines a finite list of internal transitions. By concatenating those lists with the $k$ transitions \[
\sigma_{j-1}
\longmapsto
V(\sigma_{j-1},\eta_j),
\qquad 1\le j\le k,
\] we obtain a finite sequence of states beginning at $\iota(x)$ and ending at $\sigma_k\in\Sigma_{\rm sal}$.

The internal transitions satisfy the first alternative in \eqref{eq:calctransitions}, while the preceding $k$ transitions satisfy the second. Moreover, $x\in D_\iota$, since $\hc_0\in G_0$.

Therefore, the sequence obtained is a valid computation of $\mathscr C$ on $(f,x)$, with exactly $k$ evaluations and output $a$. Thus, \[
x\in\Omega_{\mathscr C,f},
\qquad
N_{\mathscr C}(f,x)=k\le n.
\]

We have proved both inclusions. Therefore, equality~\eqref{eq:exactdomain} holds, which proves (a).

\proofstep{3. Proof of (b): output and record.}

Fix $x\in\Omega_{T_{\mathscr C,n},f}$. By (a), there exists a valid computation of $\mathscr C$ on $(f,x)$ with $k\le n$ evaluations.

Let $\boldsymbol s$ be that computation, and retain the notation $i_1<\cdots<i_k$ for the query indices, as in Step 2 of this proof.

By the first part of the proof of (a), the execution of $T_{\mathscr C,n}$ makes exactly the queries $q_j=Q(s_{i_j})$, $1\le j\le k$, with responses $\eta_j=E_f(q_j)$.

The coincidence of the answered queries and final state $s_L$ proves~\eqref{eq:exactrecords}. Uniqueness of both executions, by Lemmas~\ref{lem:unique} and~\ref{lem:calcunique}, guarantees that the outputs and records identified are the only valid ones.

\proofstep{4. Proof of (c): number of evaluations.}

Again let $x\in\Omega_{T_{\mathscr C,n},f}$, and suppose that its record contains $k$ evaluations.

By (b), the two records coincide. By the definitions of the evaluation counts, \[
\begin{aligned}
\nu_{T_{\mathscr C,n},f}(x)
&=k\\
&=N_{\mathscr C}(f,x).
\end{aligned}
\]

Moreover, for each $r\in\Nn$, \[
\begin{aligned}
\nu^{(r)}_{T_{\mathscr C,n},f}(x)
&=\bigl|
\{j\in\{1,\ldots,k\}:r_{i_j}=r\}
\bigr|\\
&=N_{\mathscr C,r}(f,x).
\end{aligned}
\] This proves (c).

Finally, let $\mathfrak C=(\phi,\mathscr C,\mathcal U)$ be a local implementation with $v(\mathfrak C)\le n$.

By \eqref{eq:realization} and the definition of cost, \[
(f,x)\in\mathcal U
\quad\Longrightarrow\quad
\begin{cases}
x\in\Omega_{\mathscr C,f},\\
N_{\mathscr C}(f,x)
\le v(\mathfrak C)\le n.
\end{cases}
\] Applying (a), we obtain \[
(f,x)\in\mathcal U
\quad\Longrightarrow\quad
x\in\Omega_{T_{\mathscr C,n},f}.
\]

By (b) and (c), for every $(f,x)\in\mathcal U$, \[
(T_{\mathscr C,n})_f(x)
=C_f(x)=\phi(f)(x),
\] \[
\Rec_{T_{\mathscr C,n},f}(x)
=\Rec_{\mathscr C,f}(x),
\] and therefore the counting equalities~\eqref{eq:exactcounts} hold.

Therefore, all conclusions of the theorem hold.
\end{proof}

\subsection{A Procedure with Two Branches}

\begin{example}[Representation of a procedure with two branches]
\label{ex:exactrepresentation}

Consider again Newton's mapping in~\eqref{eq:newton-iteration}, for $(I_f,f,\alpha_f)\in\DD$ and $x\in I_f$.

Let us construct a procedure $\mathscr C$ computing $\phi(f)$ but performing two or three evaluations depending on the sign of $f(x)$.

To this end, consider the labeled states \[
c_0(x),\quad c_+(x,y),\quad
c_-(x,y),\quad c_2(x,y,w),
\quad i(x,y,z),\quad o(a),\quad b.
\] We define \[
\begin{aligned}
\Sigma_{\rm con}
&=\{c_0(x):x\in\R\}\\
&\quad\cup\{c_+(x,y),c_-(x,y):
x,y\in\R\}\\
&\quad\cup\{c_2(x,y,w):
x,y,w\in\R\},\\
\Sigma_{\rm int}
&=\{i(x,y,z):x,y,z\in\R\},\\
\Sigma_{\rm sal}
&=\{o(a):a\in\R\},\\
\Sigma_{\rm blo}&=\{b\},
\end{aligned}
\] where the different families represent disjoint states. Let $\Sigma$ be their union, $D_\iota=\R$, $D_V=\Sigma_{\rm con}\times\R$, and \[
\iota(x)=c_0(x).
\]

The corresponding queries and transitions are defined by \[
\begin{aligned}
Q(c_0(x))&=(x,0),\\
V(c_0(x),y)&=
\begin{cases}
c_+(x,y),&y>0,\\
c_-(x,y),&y\le0,
\end{cases}\\[1mm]
Q(c_+(x,y))&=(x,1),\\
V(c_+(x,y),z)&=i(x,y,z),\\[1mm]
Q(c_-(x,y))&=(x,2),\\
V(c_-(x,y),w)&=c_2(x,y,w),\\[1mm]
Q(c_2(x,y,w))&=(x,1),\\
V(c_2(x,y,w),z)&=i(x,y,z).
\end{aligned}
\] Finally, \[
U(i(x,y,z))=
\begin{cases}
o(x-y/z),&z\ne0,\\
b,&z=0,
\end{cases}
\qquad A(o(a))=a.
\]

These mappings define a computational procedure in the sense of Definition~\ref{def:calc}.

The preceding procedure may be written as the following iteration: \[
\begin{cases}
x_0\in I_f,\\[1mm]
y_n=f(x_n),\\[2mm]
\begin{cases}
z_n=f'(x_n),&y_n>0,\\
w_n=f''(x_n),\quad z_n=f'(x_n),
&y_n\le0,
\end{cases}\\[3mm]
x_{n+1}=x_n-\dfrac{y_n}{z_n},
\end{cases}
\] whenever $x_n\in I_f$ and $f'(x_n)\ne0$.

Note that, in both cases, the output corresponds to Newton's iteration~\eqref{eq:newton-iteration}, applied at $x_n$. However, the number of evaluations depends on the sign of $f(x_n)$. In particular, the evaluation $f''(x_n)$ in the second branch does not enter the output.

\medskip
\noindent

\textbf{(a) Domain and budget.}

Let $f\in C^\infty(I_f)$. For $x\in I_f$ with $f'(x)\ne0$, the procedure makes the following queries: \[
\begin{array}{c|c|c}
\text{Condition}&\text{Queries}&
N_{\mathscr C}(f,x)\\ \hline
f(x)>0&(x,0),(x,1)&2\\
f(x)\le0&(x,0),(x,2),(x,1)&3
\end{array}
\] In both cases, $C_f(x)$ coincides with Newton's mapping~\eqref{eq:newton-iteration}.

If $x\notin I_f$, the first query is unavailable. If $x\in I_f$ and $f'(x)=0$, the procedure reaches the blocking state $b$. Therefore, \[
\Omega_{\mathscr C,f}
=\{x\in I_f:f'(x)\ne0\}.
\]

Now consider the methods $T_{\mathscr C,2}$ and $T_{\mathscr C,3}$ constructed in Theorem~\ref{thm:exactrepresentation}.

By \eqref{eq:exactdomain}, \[
\Omega_{T_{\mathscr C,2},f}
=\{x\in I_f:
f'(x)\ne0,\ f(x)>0\},
\] whereas \[
\Omega_{T_{\mathscr C,3},f}
=\{x\in I_f:f'(x)\ne0\}.
\]

Thus, the method with budget two preserves only executions using at most two evaluations, whereas the method with budget three preserves all valid executions of $\mathscr C$.

\medskip
\noindent\textbf{(b) Outputs and records.}

Take \[
I_f=(-1,1),\qquad
f(u)=u,\qquad \alpha_f=0.
\] Then $(I_f,f,0)\in\DD$.

For $x=\frac12$, we have \[
f(x)=\frac12>0,
\] and the record of the procedure is \[
\Rec_{\mathscr C,f}\left(\frac12\right)
=
\left(
\left(\frac12,0,\frac12\right),
\left(\frac12,1,1\right)
\right).
\] The output is zero, as may be verified by substituting $x=\frac12$, $f(x)=\frac12$, and $f'(x)=1$ into~\eqref{eq:newton-iteration}.

Since two evaluations are performed, equalities~\eqref{eq:exactrecords} guarantee that $T_{\mathscr C,2}$ preserves the record and output zero at $x=\frac12$.

Now consider $x=-\frac12$. Since \[
f(x)=-\frac12<0,
\] the queries are \[
\left(-\frac12,0\right),\quad
\left(-\frac12,2\right),\quad
\left(-\frac12,1\right),
\] and their responses are \[
-\frac12,\qquad 0,\qquad 1.
\] Therefore, \[
\begin{aligned}
\Rec_{\mathscr C,f}
\left(-\frac12\right)
=\biggl(
&\left(-\frac12,0,-\frac12\right),\\
&\left(-\frac12,2,0\right),\\
&\left(-\frac12,1,1\right)
\biggr),
\end{aligned}
\] with output zero, by~\eqref{eq:newton-iteration} with $x=-\frac12$, $f(x)=-\frac12$, and $f'(x)=1$.

This execution uses three evaluations. Consequently, \[
-\frac12\notin
\Omega_{T_{\mathscr C,2},f},
\qquad
-\frac12\in
\Omega_{T_{\mathscr C,3},f}.
\] Moreover, by~\eqref{eq:exactrecords}, $T_{\mathscr C,3}$ preserves that record and output zero.

\medskip
\noindent\textbf{(c) Number of evaluations.}

The records in part (b) show that, for $x=\frac12$, two queries are made, one of order zero and another of order one. For $x=-\frac12$, three queries are made, one of each order $0,1,2$. In both cases, all other counts by order are zero. By~\eqref{eq:exactcounts}, these profiles coincide with those of $T_{\mathscr C,2}$ and $T_{\mathscr C,3}$, respectively.

Finally, consider the local set $\mathcal U$ defined in~\eqref{eq:newton-local-U}. Since $I_f$ is open and $\alpha_f\in I_f$, the set $\mathcal U$ satisfies \eqref{eq:localset}.

Moreover, for every $(f,x)\in\mathcal U$, \[
C_f(x)=\phi(f)(x),
\qquad
N_{\mathscr C}(f,x)\le3.
\] Therefore, $\mathfrak C=(\phi,\mathscr C,\mathcal U)$ is a local implementation and $v(\mathfrak C)=3.$ Indeed, for each $f\in\DD$ there are points of $I_f$ with $f(x)<0$, since $f$ is strictly monotone and $f(\alpha_f)=0$.

Thus, $n=3$ is a local budget for $\mathfrak C$, and $T_{\mathscr C,3}$ preserves its outputs, records, and evaluation counts throughout $\mathcal U$.
\end{example}

\subsection{Scope of the Representation}

Equality~\eqref{eq:exactdomain} preserves only valid computations within budget $n$. The partial mapping $B$ used in its proof is defined by the existence of finite stretches of internal transitions, whose detection is not assumed computable. Neither the numbers of internal operations nor the execution times are necessarily preserved.

\par\medskip
\needspace{8\baselineskip}

\section{Reverse Representation and Comparison of the Models}\label{sec:inversa}

We now consider representing a method as a procedure. The state of the procedure will store the level and the response history. In this way, the decisions of the method are reproduced by query and output states.

\begin{proposition}[Representation of a method as a procedure]
\label{prop:methodprocedure}

Let $T$ be a method with budget $d\ge1$. Then there exists a computational procedure $\mathscr C_T$, constructed solely from $T$, satisfying the following properties for every $f\in C^\infty(I_f)$:

\begin{enumerate}[label=(\alph*)]

\item The execution domains coincide: \[
\Omega_{\mathscr C_T,f}=\Omega_{T,f}.
\]

\item For every $x\in\Omega_{T,f}$, the outputs and records coincide: \[
C_f(x)=T_f(x),
\qquad
\Rec_{\mathscr C_T,f}(x)=\Rec_{T,f}(x).
\]

\item For every $x\in\Omega_{T,f}$, the total number of evaluations and the numbers of evaluations by order are preserved: \[
\begin{aligned}
N_{\mathscr C_T}(f,x)&=\nu_{T,f}(x),\\
N_{\mathscr C_T,r}(f,x)&=\nu^{(r)}_{T,f}(x),
\qquad r\in\Nn.
\end{aligned}
\]

\end{enumerate}

\end{proposition}

\begin{proof}

\proofstep{1. Construction of the procedure.}

We use the notation for the components of $T$ given in Definition~\ref{def:method} and recall that $H_k=\R^{k+1}$. We define \[
\Sigma:=\bigcup_{k=0}^d\bigl(\{k\}\times H_k\bigr),
\qquad
\Sigma_{\rm int}:=\varnothing.
\] The first coordinate of a state $(k,\hc)$ indicates the number of responses incorporated.

Consider the sets \[
\begin{aligned}
\Sigma_{\rm con}:=
\bigl\{(k,\hc)\in\Sigma:\;&k<d,\ \hc\in D_k,\\
&\Phi_k(\hc)\in\mathcal A_c\bigr\},
\end{aligned}
\] and \[
\Sigma_{\rm sal}:=
\bigl\{(k,\hc)\in\Sigma:
\hc\in D_k,\ \Phi_k(\hc)\in\mathcal A_s\bigr\}.
\] Since $\mathcal A_c\cap\mathcal A_s=\varnothing$, a single rule cannot determine both a query and an output. Therefore, $\Sigma_{\rm con}\cap\Sigma_{\rm sal}=\varnothing.$ Moreover, $\Sigma_{\rm int}=\varnothing$, so the three classes are pairwise disjoint. The remaining states belong to \[
\Sigma_{\rm blo}:=
\Sigma\setminus(\Sigma_{\rm con}\cup\Sigma_{\rm sal}).
\]

Define initialization by \[
D_\iota:=\R,
\qquad
\iota(x):=(0,(x)).
\] Since $(x)\in H_0$, we have $\iota(x)\in\Sigma$.

For $(k,\hc)\in\Sigma_{\rm con}$, there exists a unique $q\in\QQ$ such that $\Phi_k(\hc)=(\ct,q).$ Define $Q(k,\hc):=q$. Similarly, for $(k,\hc)\in\Sigma_{\rm sal}$, there exists a unique $a\in\R$ such that \[
\Phi_k(\hc)=(\st,a),
\] and we define $A(k,\hc):=a$. Thus, the mappings \[
Q:\Sigma_{\rm con}\longrightarrow\QQ,
\qquad
A:\Sigma_{\rm sal}\longrightarrow\R.
\] are well defined.

To incorporate a response, take \[
D_V:=\Sigma_{\rm con}\times\R,
\qquad
V((k,\hc),\eta):=(k+1,\hc\Vert\eta).
\] If $((k,\hc),\eta)\in D_V$, then \[
k<d,
\qquad
\hc\in H_k,
\qquad
\hc\Vert\eta\in H_{k+1}.
\] Therefore, \[
V((k,\hc),\eta)\in\{k+1\}\times H_{k+1}
\subseteq\Sigma,
\] and $V:D_V\to\Sigma$ is well defined.

Finally, let $U:\varnothing\to\Sigma$ be the empty mapping. The preceding sets and mappings determine a procedure $\mathscr C_T$ in the sense of Definition~\ref{def:calc}. All its components depend only on $T$.

\proofstep{2. Proof of (a): equality of domains.}

Fix $f\in C^\infty(I_f)$. We first prove that \[
\Omega_{T,f}\subseteq\Omega_{\mathscr C_T,f}.
\] Let $x\in\Omega_{T,f}$ and consider its valid execution $(\Rec_k,a)$, where \[
\Rec_k=\bigl((z_j,r_j,\eta_j)\bigr)_{j=1}^k,
\qquad 0\le k\le d.
\] We define \[
q_j:=(z_j,r_j),\qquad
\hc_0:=(x),\qquad
\hc_j:=(x,\eta_1,\ldots,\eta_j).
\] By \eqref{eq:prefix} and \eqref{eq:exit}, \[
\begin{gathered}
\hc_j\in D_j\quad(0\le j\le k),\\
\Phi_j(\hc_j)=(\ct,q_{j+1}),\quad
q_{j+1}\in\QQ_f,\quad
\eta_{j+1}=E_f(q_{j+1})\quad(0\le j<k),\\
\Phi_k(\hc_k)=(\st,a).
\end{gathered}
\]

Consider the states \[
s_j:=(j,\hc_j),\qquad 0\le j\le k.
\] Then \[
s_0=\iota(x),\qquad
s_j\in\Sigma_{\rm con}\quad(0\le j<k),
\qquad s_k\in\Sigma_{\rm sal}.
\] Moreover, for $0\le j<k$, \[
Q(s_j)=q_{j+1}\in\QQ_f,
\qquad
(s_j,E_f(Q(s_j)))\in D_V,
\] and \[
\begin{aligned}
V(s_j,E_f(Q(s_j)))
&=V((j,\hc_j),\eta_{j+1})\\
&=(j+1,\hc_j\Vert\eta_{j+1})\\
&=(j+1,\hc_{j+1})=s_{j+1}.
\end{aligned}
\] Therefore, $(s_0,\ldots,s_k)$ satisfies \eqref{eq:calcstart} and \eqref{eq:calctransitions}. Its output is $A(s_k)=a.$ This also includes the case $k=0$: then $s_0\in\Sigma_{\rm sal}$ and there are no transitions. We conclude that $x\in\Omega_{\mathscr C_T,f}$.

We now prove the reverse inclusion. Let $x\in\Omega_{\mathscr C_T,f}$ and consider its valid computation $\boldsymbol s=(s_0,\ldots,s_L)$. Since $\Sigma_{\rm int}=\varnothing$, all its transitions are query transitions. Moreover, \[
s_0=(0,(x)),
\] and each application of $V$ increases the first coordinate by one and appends a response to the history. By induction, there exist histories $\hc_j\in H_j$ such that \[
s_j=(j,\hc_j),\qquad 0\le j\le L.
\] Since $s_L\in\Sigma$, necessarily $L\le d$.

For $0\le j<L$, define \[
q_{j+1}:=Q(s_j)=(z_{j+1},r_{j+1}),
\qquad
\eta_{j+1}:=E_f(q_{j+1}).
\] By the validity of the computation and the definition of $Q$, \[
\hc_j\in D_j,\qquad
\Phi_j(\hc_j)=(\ct,q_{j+1}),\qquad
q_{j+1}\in\QQ_f.
\] Moreover, \[
s_{j+1}=V(s_j,\eta_{j+1})
=(j+1,\hc_j\Vert\eta_{j+1}).
\] Since $\hc_0=(x)$, we obtain by induction \[
\hc_j=(x,\eta_1,\ldots,\eta_j),
\qquad 0\le j\le L.
\] Finally, $s_L\in\Sigma_{\rm sal}$ implies \[
\hc_L\in D_L,
\qquad
\Phi_L(\hc_L)=(\st,A(s_L)).
\] Thus, \[
\Rec_L:=\bigl((z_j,r_j,\eta_j)\bigr)_{j=1}^L
\] satisfies \eqref{eq:prefix} and \eqref{eq:exit}, with output $a=A(s_L)$. Therefore, $x\in\Omega_{T,f}$.

We have proved both inclusions and hence the equality of domains in (a).

\proofstep{3. Proof of (b) and (c).}

Let $x$ be a point of the common domain. The construction in the preceding step transforms the record of $T$ into a computation of $\mathscr C_T$ that makes the same queries, with the same responses and in the same order.

By Lemmas~\ref{lem:unique} and \ref{lem:calcunique}, the valid executions and computations are unique. The equalities of records and outputs in (b) follow.

Write the common record using the notation of~\eqref{eq:generic-T-register}. By the definitions of the counts, \[
N_{\mathscr C_T}(f,x)=k=\nu_{T,f}(x).
\] Likewise, for each $r\in\Nn$, \[
\begin{aligned}
N_{\mathscr C_T,r}(f,x)
&=\bigl|\{j\in\{1,\ldots,k\}:r_j=r\}\bigr|\\
&=\nu^{(r)}_{T,f}(x).
\end{aligned}
\] These equalities also hold for $k=0$, since both records are empty and all counts are zero. This proves (c).

\end{proof}

\subsection{Joint Interpretation of the Two Representations}

Let $T$ be a method with budget $d$. Proposition~\ref{prop:methodprocedure} represents it by $\mathscr C_T$ with the same domain, record, output, and counts; in particular, $N_{\mathscr C_T}(f,x)\le d$ on its domain. By Theorem~\ref{thm:exactrepresentation}, the representation of $\mathscr C_T$ with budget $d$ satisfies \[
\Omega_{T_{\mathscr C_T,d},f}=\Omega_{T,f}.
\] Conversely, when a procedure $\mathscr C$ is first represented with budget $n$ and then as a procedure, the truncated domain in~\eqref{eq:exactdomain} is preserved exactly, together with records, outputs, and counts. These compositions do not identify internal states and are not inverse as tuples of mappings.

\par\medskip
\needspace{8\baselineskip}

\section{Reuse of Responses}\label{sec:reutilizacion}

We study how to store a response during a computation to avoid repeated acquisitions, while preserving validity and output.

\begin{lemma}[Reuse of responses without new evaluations]
\label{lem:cache}

Let $\mathscr C$ be a computational procedure. Then there exists another procedure $\mathscr C^{\rm mem}$, constructed solely from $\mathscr C$, such that, for every $f\in C^\infty(I_f)$, \[
\Omega_{\mathscr C^{\rm mem},f}=\Omega_{\mathscr C,f}.
\] Moreover, for every $x$ in this domain, \[
C_f^{\rm mem}(x)=C_f(x).
\]

If the record of the original computation is \[
\Rec_{\mathscr C,f}(x)
=\bigl((z_j,r_j,\eta_j)\bigr)_{j=1}^k,
\] then \[
N_{\mathscr C^{\rm mem}}(f,x)
=\bigl|\{(z_j,r_j):1\le j\le k\}\bigr|,
\] and, for each $r\in\Nn$, \[
\begin{aligned}
N_{\mathscr C^{\rm mem},r}(f,x)
&=\bigl|\{z_j:1\le j\le k,\ r_j=r\}\bigr|\\
&\le N_{\mathscr C,r}(f,x).
\end{aligned}
\]

In particular, the new procedure performs only one evaluation for each distinct query $(z,r)$ and reuses its response at subsequent occurrences. The list of stored responses is empty at the start of each new computation.

\end{lemma}

\begin{proof}

\proofstep{1. Lists of stored queries and responses.}

Let \[
\mathscr C=
(\Sigma,\Sigma_{\rm int},\Sigma_{\rm con},
\Sigma_{\rm sal},D_\iota,D_V,\iota,U,Q,V,A).
\] Consider the set \[
\mathcal M:=\bigcup_{h\in\Nn}(\QQ\times\R)^h.
\] Its elements are finite lists of query--response pairs; for $h=0$, we have the empty list.

For \[
M=((q_j,\eta_j))_{j=1}^h\in\mathcal M
\quad\text{and}\quad q\in\QQ,
\] define \[
I_M(q):=\{j\in\{1,\ldots,h\}:q_j=q\}.
\] If $I_M(q)\ne\varnothing$, this set is finite and has a least element. In that case, define \[
e_M(q):=\eta_{\min I_M(q)}.
\] Thus, $e_M(q)$ is the first stored response to query $q$. The notation $M\Vert(q,\eta)$ means that we append the pair $(q,\eta)$ as a new element of the list.

\proofstep{2. States and mappings of the new procedure.}

We define \[
\Sigma':=\Sigma\times\mathcal M.
\] In a state $(s,M)$, the first coordinate represents the state of the original procedure, whereas the second contains the responses stored during the current computation.

We take \[
\Sigma'_{\rm sal}:=\Sigma_{\rm sal}\times\mathcal M,
\] \[
\Sigma'_{\rm con}:=
\{(s,M)\in\Sigma':
s\in\Sigma_{\rm con},\ I_M(Q(s))=\varnothing\},
\] and \[
\begin{aligned}
\Sigma'_{\rm int}:={}&(\Sigma_{\rm int}\times\mathcal M)\\
&\cup\left\{(s,M)\in\Sigma':
\begin{array}{l}
s\in\Sigma_{\rm con},\\
I_M(Q(s))\ne\varnothing,\\
(s,e_M(Q(s)))\in D_V
\end{array}\right\}.
\end{aligned}
\]

Note that a query of the original procedure becomes an internal transition when its response has already been stored and can be incorporated by $V$.

The sets are disjoint: their first coordinates belong to different classes, except for the original query states, which are separated by the mutually exclusive conditions $I_M(Q(s))=\varnothing$ and $I_M(Q(s))\ne\varnothing$. We define blocking states by \[
\Sigma'_{\rm blo}:=
\Sigma'\setminus
(\Sigma'_{\rm int}\cup\Sigma'_{\rm con}\cup\Sigma'_{\rm sal}).
\]

For initialization, take \[
D'_\iota:=D_\iota,
\qquad
\iota'(x):=(\iota(x),()).
\] The query and output mappings are given by \[
Q'(s,M):=Q(s)
\quad((s,M)\in\Sigma'_{\rm con}),
\] \[
A'(s,M):=A(s)
\quad((s,M)\in\Sigma'_{\rm sal}).
\]

To incorporate a new response, define \[
D'_V:=
\{((s,M),\eta)\in\Sigma'_{\rm con}\times\R:
(s,\eta)\in D_V\}
\] and \[
V'((s,M),\eta):=
\bigl(V(s,\eta),M\Vert(Q(s),\eta)\bigr).
\] If $((s,M),\eta)\in D'_V$, both coordinates of $V'((s,M),\eta)$ belong, respectively, to $\Sigma$ and $\mathcal M$; hence $V'$ takes values in $\Sigma'$.

Finally, for $(s,M)\in\Sigma'_{\rm int}$, define \[
U'(s,M):=
\begin{cases}
(U(s),M),&s\in\Sigma_{\rm int},\\[1mm]
(V(s,e_M(Q(s))),M),&s\in\Sigma_{\rm con}.
\end{cases}
\] In the second case, the definition of $\Sigma'_{\rm int}$ guarantees that $e_M(Q(s))$ is defined and that \[
(s,e_M(Q(s)))\in D_V.
\] Thus, both expressions are well defined and belong to $\Sigma'$.

We conclude that \[
\mathscr C^{\rm mem}=
(\Sigma',\Sigma'_{\rm int},\Sigma'_{\rm con},
\Sigma'_{\rm sal},D'_\iota,D'_V,\iota',U',Q',V',A')
\] is a computational procedure. Its construction uses only the components of $\mathscr C$.

\proofstep{3. Validity of the stored responses.}

Fix $f\in C^\infty(I_f)$. For $M=((q_j,\eta_j))_{j=1}^h$, consider the property \[
\mathcal I_f(M):\qquad
q_j\in\QQ_f,\quad\eta_j=E_f(q_j),
\qquad 1\le j\le h.
\] We show that this property holds for every list appearing in a valid computation of the new procedure.

The initial list is empty, so $\mathcal I_f(())$ holds. An internal transition does not modify the list. If a new query is made, validity of the computation requires \[
q=Q'(s,M)\in\QQ_f,
\qquad\eta=E_f(q),
\] and the new list is $M\Vert(q,\eta)$. Therefore, \[
\mathcal I_f(M)
\quad\Longrightarrow\quad
\mathcal I_f(M\Vert(q,E_f(q))).
\] We conclude by induction that the property is preserved throughout every valid computation.

In particular, if $\mathcal I_f(M)$ and $I_M(q)\ne\varnothing$, there exists $j=\min I_M(q)$ and \[
q=q_j\in\QQ_f,
\qquad
e_M(q)=\eta_j=E_f(q).
\] Thus, \begin{equation}\label{eq:cacheanswer}
\mathcal I_f(M),\quad I_M(q)\ne\varnothing
\quad\Longrightarrow\quad
q\in\QQ_f,\qquad e_M(q)=E_f(q).
\end{equation}

\proofstep{4. Preservation of domains and outputs.}

We first prove that \[
\Omega_{\mathscr C,f}
\subseteq\Omega_{\mathscr C^{\rm mem},f}.
\] Let $x\in\Omega_{\mathscr C,f}$ and consider its valid computation $(s_0,\ldots,s_L)$. We shall construct states \[
\widehat s_i=(s_i,M_i),\qquad 0\le i\le L,
\] forming a valid computation of the new procedure.

Initially, take \[
M_0:=(),\qquad
\widehat s_0=(\iota(x),())=\iota'(x).
\] Suppose that $\widehat s_i=(s_i,M_i)$ has been constructed, with $i<L$ and $\mathcal I_f(M_i)$.

If $s_i\in\Sigma_{\rm int}$, then \[
s_{i+1}=U(s_i),\qquad
\widehat s_i\in\Sigma'_{\rm int},
\] and \[
U'(\widehat s_i)=(U(s_i),M_i)=(s_{i+1},M_i).
\] We take $M_{i+1}:=M_i$.

If $s_i\in\Sigma_{\rm con}$, define \[
q:=Q(s_i),\qquad\eta:=E_f(q).
\] By validity of the original computation, \[
q\in\QQ_f,\qquad
(s_i,\eta)\in D_V,\qquad
s_{i+1}=V(s_i,\eta).
\] We distinguish two cases.

If $I_{M_i}(q)=\varnothing$, we have \[
\widehat s_i\in\Sigma'_{\rm con},\qquad
Q'(\widehat s_i)=q,\qquad
(\widehat s_i,\eta)\in D'_V.
\] Therefore, \[
V'(\widehat s_i,\eta)
=(s_{i+1},M_i\Vert(q,\eta)),
\] and we take $M_{i+1}:=M_i\Vert(q,\eta)$. The property $\mathcal I_f(M_{i+1})$ is preserved, since $\eta=E_f(q)$.

If $I_{M_i}(q)\ne\varnothing$, by \eqref{eq:cacheanswer}, \[
e_{M_i}(q)=E_f(q)=\eta.
\] Since $(s_i,\eta)\in D_V$, we obtain \[
\widehat s_i\in\Sigma'_{\rm int}
\] and \[
\begin{aligned}
U'(\widehat s_i)
&=(V(s_i,e_{M_i}(q)),M_i)\\
&=(V(s_i,\eta),M_i)=(s_{i+1},M_i).
\end{aligned}
\] In this case, take $M_{i+1}:=M_i$.

Thus, in every case, there is a valid transition with first coordinate $s_{i+1}$, and $\mathcal I_f(M_{i+1})$ is preserved. By induction, we obtain the finite list \[
(\widehat s_0,\ldots,\widehat s_L).
\] Moreover, \[
s_L\in\Sigma_{\rm sal}
\quad\Longrightarrow\quad
\widehat s_L=(s_L,M_L)\in\Sigma'_{\rm sal},
\] and \[
A'(\widehat s_L)=A(s_L).
\] Therefore, \[
x\in\Omega_{\mathscr C^{\rm mem},f},
\qquad C_f^{\rm mem}(x)=C_f(x).
\]

We now prove the reverse inclusion. Let $x\in\Omega_{\mathscr C^{\rm mem},f}$ and consider its valid computation \[
(\widehat s_0,\ldots,\widehat s_L),
\qquad\widehat s_i=(s_i,M_i).
\] By initialization, \[
x\in D_\iota,\qquad s_0=\iota(x),\qquad M_0=().
\] Moreover, by Step 3, all lists $M_i$ satisfy $\mathcal I_f(M_i)$. We show that $(s_0,\ldots,s_L)$ is a valid computation of $\mathscr C$.

Fix $0\le i<L$. If $\widehat s_i\in\Sigma'_{\rm con}$, the validity of the transition and the definition of $V'$ imply \[
\begin{gathered}
s_i\in\Sigma_{\rm con},\qquad Q(s_i)\in\QQ_f,\\
(s_i,E_f(Q(s_i)))\in D_V,\\
s_{i+1}=V(s_i,E_f(Q(s_i))).
\end{gathered}
\] Therefore, this is a valid query transition of the original procedure.

If $\widehat s_i\in\Sigma'_{\rm int}$ and $s_i\in\Sigma_{\rm int}$, then \[
s_{i+1}=U(s_i),
\] which is a valid internal transition.

Finally, if $\widehat s_i\in\Sigma'_{\rm int}$ and $s_i\in\Sigma_{\rm con}$, then, by construction, \[
I_{M_i}(Q(s_i))\ne\varnothing,
\qquad (s_i,e_{M_i}(Q(s_i)))\in D_V.
\] By \eqref{eq:cacheanswer}, \[
Q(s_i)\in\QQ_f,
\qquad e_{M_i}(Q(s_i))=E_f(Q(s_i)).
\] Therefore, \[
s_{i+1}=V(s_i,E_f(Q(s_i))),
\] and this is also a valid query transition of the original procedure.

We have verified all transitions. Moreover, \[
\widehat s_L\in\Sigma'_{\rm sal}
\quad\Longrightarrow\quad s_L\in\Sigma_{\rm sal}.
\] Thus, $(s_0,\ldots,s_L)$ is a valid computation of $\mathscr C$ on $(f,x)$, with output \[
A(s_L)=A'(\widehat s_L).
\] We conclude that the equality of domains and the preservation of outputs stated in the lemma hold for every $x$ in the common domain.

\proofstep{5. Number of evaluations and reinitialization of the list.}

Let $x\in\Omega_{\mathscr C,f}$ and write \[
\Rec_{\mathscr C,f}(x)
=\bigl((z_j,r_j,\eta_j)\bigr)_{j=1}^k,
\qquad q_j:=(z_j,r_j).
\] Before the first query, the stored list is empty. Each new query appends its query--response pair; repeated queries and internal transitions do not modify the list. By induction on $j$, immediately before the original query $q_j$, the stored queries are precisely the distinct elements of \[
\{q_1,\ldots,q_{j-1}\}.
\]

Therefore, the new procedure makes an evaluation at that transition if and only if \[
q_j\notin\{q_1,\ldots,q_{j-1}\}.
\] At other occurrences, it uses the stored response by means of $U'$, without a new evaluation. Thus, exactly one evaluation is performed for each element of the set $\{q_1,\ldots,q_k\}$, proving the total-count formula in the statement.

For a fixed order $r\in\Nn$, two queries $(z,r)$ and $(z',r)$ coincide if and only if $z=z'$. Therefore, the number of distinct nodes does not exceed the number of occurrences of order $r$. This establishes the inequality by orders stated in the lemma. In particular, \[
N_{\mathscr C^{\rm mem}}(f,x)\le N_{\mathscr C}(f,x).
\] If $k=0$, the preceding sets are empty and all equalities remain valid.

Each original transition corresponds to a transition of the new procedure, although some queries become internal transitions. By initialization $\iota'(x)=(\iota(x),())$, the memory starts empty and retains no responses from previous computations.

\end{proof}

\subsection{Effective Acquisitions and Absence of Memory between Iterations}

The effective record of the procedure with memory may omit repeated queries from the original record, even though it preserves the output and domain. The memory is reset at each computation, in accordance with the convention that no information is inherited between iterations.

\par\medskip
\needspace{8\baselineskip}

\section{Staged Representations}\label{sec:etapas}

We shall represent by procedures implementations whose data are acquired in successive blocks.

\label{subsec:stages-pruned}

\begin{definition}[Staged implementation]
\label{def:stages}

\begin{enumerate}

\item[(a)] \textbf{Implementation data.}

Let $\ell\in\N$. For each $0\le j<\ell$, consider a finite set \[
S_j=\{r_{j,1}<\cdots<r_{j,m_j}\}
\subseteq\Nn,
\qquad m_j\in\Nn,
\] and define \[
a_j:=\sum_{i=0}^{j-1}m_i,
\qquad a_0=0.
\]

Let \[
W_j\subseteq\R^{1+a_j+m_j}
\] and \[
\bar u_{j+1}:W_j\longrightarrow\R,
\qquad 0\le j<\ell.
\]

We denote by \[
\mathfrak R=
\bigl(
\ell,(S_j)_{j=0}^{\ell-1},
(W_j)_{j=0}^{\ell-1},
(\bar u_{j+1})_{j=0}^{\ell-1}
\bigr)
\] the tuple formed by these objects.

We allow $m_j=0$, in which case $S_j=\varnothing$, and adopt the convention \[
\R^0=\{()\}.
\]

The sets $W_j$ and mappings $\bar u_{j+1}$ are fixed independently of the function on which the computation is performed.

\item[(b)] \textbf{Valid evaluation.}

Let $f\in C^\infty(I_f)$ and $x\in\R$. We say that $\mathfrak R$ admits a valid evaluation on $(f,x)$ if there exist real numbers \[
z_0,\ldots,z_\ell
\] and blocks \[
B_j\in\R^{m_j},
\qquad 0\le j<\ell,
\] satisfying \begin{equation}\label{eq:reducedstages}
\begin{gathered}
z_0=x,\\[1mm]
m_j>0\ \Longrightarrow\ z_j\in I_f,\\[1mm]
B_j=
\begin{cases}
\bigl(
f^{(r_{j,1})}(z_j),\ldots,
f^{(r_{j,m_j})}(z_j)
\bigr),
&m_j>0,\\[1mm]
(),&m_j=0,
\end{cases}\\[2mm]
(x;B_0,\ldots,B_j)\in W_j,\\[1mm]
z_{j+1}
=\bar u_{j+1}(x;B_0,\ldots,B_j),
\qquad 0\le j<\ell.
\end{gathered}
\end{equation}

Here $(x;B_0,\ldots,B_j)$ denotes the vector obtained by concatenating $x$ and the coordinates of the blocks $B_0,\ldots,B_j$. In particular, \[
(x;B_0,\ldots,B_j)
\in\R^{1+a_j+m_j}.
\]

If $m_j=0$, no evaluation is made at $z_j$ and the corresponding block is empty.

The output of the valid evaluation is $z_\ell$.

\item[(c)] \textbf{Implementation of a mapping.}

Let $\phi$ be an assignment as in Definition~\ref{def:realizedcost}, and let $\mathcal U$ be a local input set.

We say that $\mathfrak R$ is a staged implementation of $\phi$ on $\mathcal U$ if, for every $(f,x)\in\mathcal U$, there exists a valid evaluation of $\mathfrak R$ on $(f,x)$ and its output satisfies \[
z_\ell=\phi(f)(x).
\]

\end{enumerate}
\end{definition}

\begin{proposition}[Staged construction and exact count]
\label{prop:stages}

Let $\mathfrak R$ be a tuple as in Definition~\ref{def:stages}. Then there exists a computational procedure $\mathscr C_{\mathfrak R}$, constructed solely from $\mathfrak R$, which satisfies the following properties for every $f\in C^\infty(I_f)$.

\begin{enumerate}

\item[(a)] \textbf{Evaluations and outputs.}

For every $x\in\R$, \[
x\in\Omega_{\mathscr C_{\mathfrak R},f}
\quad\Longleftrightarrow\quad
\mathfrak R\text{ admits a valid evaluation
on }(f,x).
\]

In this case, for each $0\le j<\ell$, the procedure makes the queries \[
(z_j,r_{j,1}),\ldots,(z_j,r_{j,m_j}),
\] in the indicated order, and produces the output \[
C_f(x)=z_\ell.
\]

\item[(b)] \textbf{Number of evaluations and fixed profile.}

For each $r\in\Nn$, define \[
c_r:=
\bigl|\{j\in\{0,\ldots,\ell-1\}:
r\in S_j\}\bigr|.
\]

Then, for every $x\in\Omega_{\mathscr C_{\mathfrak R},f}$, \[
N_{\mathscr C_{\mathfrak R},r}(f,x)=c_r,
\qquad r\in\Nn,
\] and \[
N_{\mathscr C_{\mathfrak R}}(f,x)
=\sum_{j=0}^{\ell-1}m_j
=\sum_{r\in\Nn}c_r.
\]

In particular, all valid computations have the same profile $(c_r)_{r\in\Nn}$ and use exactly \[
N:=\sum_{j=0}^{\ell-1}|S_j|
\] evaluations.

\end{enumerate}

Moreover, if $\mathfrak R$ is a staged implementation of $\phi$ on a local input set $\mathcal U$, then \[
\mathfrak C=
(\phi,\mathscr C_{\mathfrak R},\mathcal U)
\] is a local implementation of $\phi$, with \[
v(\mathfrak C)=N,
\qquad
v_r(\mathfrak C)=c_r,
\quad r\in\Nn.
\]

\end{proposition}

\begin{proof}
\proofstep{1. Construction of the procedure.}

Fix a tag $\mathrm{o}$ distinct from the integers and define the set of working states \[
\begin{aligned}
\Sigma_{\rm trab}:=
\bigl\{
(j,t,x,\boldsymbol B,z,b):\;&
0\le j<\ell,\quad 0\le t\le m_j,\\
&x,z\in\R,\quad
\boldsymbol B\in
\prod_{i=0}^{j-1}\R^{m_i},\\
&b\in\R^t
\bigr\}.
\end{aligned}
\] For $j=0$, the preceding product contains only the empty list.

Also consider the output states \[
\Sigma_{\rm sal}:=
\{(\mathrm{o},a):a\in\R\},
\] and take \[
\Sigma:=\Sigma_{\rm trab}
\cup\Sigma_{\rm sal}.
\]

For $s=(j,t,x,\boldsymbol B,z,b)
\in\Sigma_{\rm trab}$, define \[
\Sigma_{\rm con}:=
\{s\in\Sigma_{\rm trab}:t<m_j\},
\] and \[
\Sigma_{\rm int}:=
\left\{
s\in\Sigma_{\rm trab}:
\begin{array}{l}
t=m_j,\\
(x;\boldsymbol B,b)\in W_j
\end{array}
\right\}.
\]

The state classes are disjoint: query states satisfy $t<m_j$, internal states satisfy $t=m_j$, and output states carry the distinct tag $\mathrm{o}$.

The remaining states belong to \[
\Sigma_{\rm blo}:=
\Sigma\setminus
(\Sigma_{\rm int}\cup
\Sigma_{\rm con}\cup
\Sigma_{\rm sal}).
\]

We now define the mappings of the procedure.

We take \[
D_\iota:=\R,
\qquad
\iota(x):=(0,0,x,(),x,()).
\] Since $\ell\ge1$, we have $\iota(x)\in\Sigma_{\rm trab} \subseteq\Sigma.$

For each $s=(j,t,x,\boldsymbol B,z,b)
\in\Sigma_{\rm con}$, define $Q(s):=(z,r_{j,t+1}).$ Since $0\le t<m_j$, the index $t+1$ belongs to $\{1,\ldots,m_j\}$; therefore $Q$ is well defined and takes values in $\QQ$.

To incorporate a response, take \[
D_V:=\Sigma_{\rm con}\times\R
\] and define \[
V(s,\eta):=
(j,t+1,x,\boldsymbol B,z,b\Vert\eta).
\] Since $t+1\le m_j$ and $b\Vert\eta\in\R^{t+1}$, we have $V(s,\eta)\in\Sigma_{\rm trab}.$ Thus, $V:D_V\longrightarrow\Sigma$ is well defined.

Now consider $s=(j,m_j,x,\boldsymbol B,z,b)
\in\Sigma_{\rm int}$. By the definition of this set, \[
(x;\boldsymbol B,b)\in W_j.
\] Consequently, we may define \[
z^+:=
\bar u_{j+1}(x;\boldsymbol B,b)
\in\R.
\]

Take \[
U(s):=
\begin{cases}
(j+1,0,x,\boldsymbol B\Vert b,z^+,()),
&j+1<\ell,\\[1mm]
(\mathrm{o},z^+),
&j+1=\ell.
\end{cases}
\]

In the first case, \[
\boldsymbol B\Vert b
\in\prod_{i=0}^{j}\R^{m_i},
\] so $U(s)\in\Sigma_{\rm trab}$. In the second, $U(s)\in\Sigma_{\rm sal}$.

Therefore, $U:\Sigma_{\rm int}\longrightarrow\Sigma$ is well defined.

Finally, define \[
A:\Sigma_{\rm sal}\longrightarrow\R,
\qquad A(\mathrm{o},a):=a.
\]

Thus, all the preceding sets and mappings determine a computational procedure $\mathscr C_{\mathfrak R}$ in the sense of Definition~\ref{def:calc}.

Note that its components depend only on $\mathfrak R$ and not on the function $f$ to which it is applied.

\proofstep{2. Proof of (a): evaluations and outputs.}

Fix $f\in C^\infty(I_f)$ and $x\in\R$.

First suppose that $\mathfrak R$ admits a valid evaluation on $(f,x)$. Let \[
z_0,\ldots,z_\ell,
\qquad
B_0,\ldots,B_{\ell-1}
\] be the objects satisfying \eqref{eq:reducedstages}.

For each $0\le j<\ell$ and $0\le t\le m_j$, define \[
b_{j,t}:=
\begin{cases}
(),&t=0,\\[1mm]
\bigl(
f^{(r_{j,1})}(z_j),\ldots,
f^{(r_{j,t})}(z_j)
\bigr),&t\ge1.
\end{cases}
\] In particular, \[
b_{j,m_j}=B_j.
\]

Consider the states \[
s_{j,t}:=
\bigl(
j,t,x,(B_0,\ldots,B_{j-1}),
z_j,b_{j,t}
\bigr).
\]

Since $z_0=x$, the first state $s_{0,0}$ coincides with initialization $\iota(x)$ defined in Step 1.

We show that, for each stage $j$, the states \[
s_{j,0},s_{j,1},\ldots,s_{j,m_j}
\] determine the query transitions corresponding to the block $B_j$.

Indeed, if $0\le t<m_j$, then by the definition of $\Sigma_{\rm con}$, \[
s_{j,t}\in\Sigma_{\rm con},
\] and \[
Q(s_{j,t})=(z_j,r_{j,t+1}).
\]

Since $m_j>0$, validity of \eqref{eq:reducedstages} implies $z_j\in I_f$. Therefore, \[
Q(s_{j,t})\in\QQ_f,
\qquad\text{and}\qquad
E_f(Q(s_{j,t}))
=f^{(r_{j,t+1})}(z_j).
\]

By the definition of $V$, \[
\begin{aligned}
&V\bigl(s_{j,t},
E_f(Q(s_{j,t}))\bigr)\\
&\quad=
\bigl(
j,t+1,x,(B_0,\ldots,B_{j-1}),
z_j,\\
&\hspace{43mm}
b_{j,t}\Vert
f^{(r_{j,t+1})}(z_j)
\bigr)\\
&\quad=s_{j,t+1}.
\end{aligned}
\]

Thus, by induction on $t$, we obtain the $m_j$ queries of stage $j$, in the indicated order, and reach state $s_{j,m_j}$.

Since \[
b_{j,m_j}=B_j
\] and, by \eqref{eq:reducedstages}, \[
(x;B_0,\ldots,B_j)\in W_j,
\] we have $s_{j,m_j}\in\Sigma_{\rm int}.$

Moreover, \[
z^+
=\bar u_{j+1}(x;B_0,\ldots,B_j)
=z_{j+1}.
\] Consequently, \[
U(s_{j,m_j})=
\begin{cases}
s_{j+1,0},&j+1<\ell,\\[1mm]
(\mathrm{o},z_\ell),&j+1=\ell.
\end{cases}
\]

Note that these equalities also hold when $m_j=0$. In that case, \[
s_{j,0}=s_{j,m_j},
\qquad B_j=(),
\] and the stage consists of a single internal transition.

By induction on $j=0,\ldots,\ell-1$, we conclude that there exists a valid computation of $\mathscr C_{\mathfrak R}$ on $(f,x)$, whose final state is $s_L=(\mathrm{o},z_\ell).$ Therefore, \[
x\in\Omega_{\mathscr C_{\mathfrak R},f},
\qquad
C_f(x)=A(s_L)=z_\ell.
\]

We now prove the converse implication.

Suppose that \[
x\in\Omega_{\mathscr C_{\mathfrak R},f}
\] and let \[
(s_0,\ldots,s_L)
\] be its unique valid computation.

By the definition of $\iota$, \[
s_0=(0,0,x,(),x,()).
\]

Note that, in each working state, transitions are determined by the following alternatives: \[
\begin{cases}
t<m_j
&\Longrightarrow
\text{a query transition},\\
t=m_j,\ (x;\boldsymbol B,b)\in W_j
&\Longrightarrow
\text{an internal transition}.
\end{cases}
\] In any other case, the state belongs to $\Sigma_{\rm blo}$ and cannot be part of a valid computation.

We show by induction on $j$ that the computation determines the objects in \eqref{eq:reducedstages}.

For $j=0$, we have $z_0=x$. Suppose that we have reached the beginning of a stage $j$, with state \[
s_{j,0}
=\bigl(j,0,x,(B_0,\ldots,B_{j-1}),
z_j,()\bigr).
\]

If $m_j>0$, the only possible transitions before leaving the stage are \[
Q(s_{j,t})=(z_j,r_{j,t+1}),
\qquad 0\le t<m_j.
\] Since the computation is valid, all these queries belong to $\QQ_f$. Therefore, $z_j\in I_f.$ Moreover, their responses are \[
f^{(r_{j,t+1})}(z_j),
\qquad 0\le t<m_j.
\] By the definition of $V$, after completing these queries we obtain exactly the block \[
B_j=
\bigl(
f^{(r_{j,1})}(z_j),\ldots,
f^{(r_{j,m_j})}(z_j)
\bigr).
\]

If $m_j=0$, there are no query transitions in this stage and \[
B_j=().
\]

In both cases, the computation reaches a state with $t=m_j$. Since it must continue to an output state, that state necessarily belongs to $\Sigma_{\rm int}$. Thus, $(x;B_0,\ldots,B_j)\in W_j.$

By the definition of $U$, the next node coincides with that specified in~\eqref{eq:reducedstages}.

If $j+1<\ell$, the computation continues with state $s_{j+1,0}$. If $j+1=\ell$, it ends at \[
(\mathrm{o},z_\ell).
\]

We conclude by induction that the nodes and blocks obtained satisfy all conditions in \eqref{eq:reducedstages}. Therefore, $\mathfrak R$ admits a valid evaluation on $(f,x)$.

We have proved both implications and hence the equivalence of domains stated in part (a) of the proposition.

Moreover, in both directions the queries are made in the same order and the output is $z_\ell$, which proves (a).

\proofstep{3. Proof of (b): number of evaluations and fixed profile.}

Let $x\in\Omega_{\mathscr C_{\mathfrak R},f}$.

By (a), the computation performs at each stage exactly the queries indicated in the statement. Therefore, \[
N_{\mathscr C_{\mathfrak R}}(f,x)
=\sum_{j=0}^{\ell-1}m_j
=:N.
\]

Moreover, each stage contains $m_j$ query transitions and one internal transition. Consequently, the length of the computation is \[
L=\sum_{j=0}^{\ell-1}(m_j+1)
=N+\ell<\infty.
\]

Now fix $r\in\Nn$. Since the elements of each $S_j$ are distinct, stage $j$ performs an evaluation of order $r$ if and only if $r\in S_j$.

Therefore, \[
\begin{aligned}
N_{\mathscr C_{\mathfrak R},r}(f,x)
&=\bigl|
\{j\in\{0,\ldots,\ell-1\}:
r\in S_j\}
\bigr|\\
&=c_r.
\end{aligned}
\]

Since the sets $S_j$ are fixed, the profile $(c_r)_{r\in\Nn}$ does not depend on $f$ or $x$.

Moreover, \[
\begin{aligned}
\sum_{r\in\Nn}c_r
&=\sum_{r\in\Nn}
\sum_{j=0}^{\ell-1}
\mathbf1_{S_j}(r)\\
&=\sum_{j=0}^{\ell-1}
\sum_{r\in S_j}1\\
&=\sum_{j=0}^{\ell-1}|S_j|\\
&=\sum_{j=0}^{\ell-1}m_j=N.
\end{aligned}
\] The interchange of sums is justified because $\ell$ is finite and all sets $S_j$ are finite.

We conclude that the counting identities in part (b) hold for every valid computation and $r\in\Nn$.

Finally, suppose that $\mathfrak R$ is a staged implementation of $\phi$ on $\mathcal U$.

By Definition~\ref{def:stages}, for every $(f,x)\in\mathcal U$ there exists a valid evaluation whose output is $\phi(f)(x)$. Applying (a), the procedure has a valid computation on that pair and produces the same output.

Therefore, the triple $\mathfrak C=(\phi,\mathscr C_{\mathfrak R},\mathcal U)$, defined in the statement, is a local implementation of $\phi$.

Moreover, by (b), \[
\begin{aligned}
v(\mathfrak C)
&=\sup_{(f,x)\in\mathcal U}
N_{\mathscr C_{\mathfrak R}}(f,x)\\
&=N,
\end{aligned}
\] and, for each $r\in\Nn$, \[
v_r(\mathfrak C)=c_r.
\]

Thus, $\mathfrak C$ has local budget $N$ and fixed profile $(c_r)_{r\in\Nn}$.
\end{proof}

\subsection{Newton's Method as a Staged Implementation}

Let us return to Newton's method, already described through queries and through states. In Definition~\ref{def:stages}, take \[
\ell=1,\qquad S_0=\{0,1\},\qquad
W_0=\{(x,a,b)\in\R^3:b\ne0\},
\] and define \[
\bar u_1(x,a,b)=x-\frac{a}{b}.
\] The block of the single stage is $B_0=(f(x),f'(x))$, available when $x\in I_f$. The condition $(x;B_0)\in W_0$ is equivalent to $f'(x)\ne0$; therefore, the implementation has Newton's domain and output, determined by $x\in I_f$, $f'(x)\ne0$, and~\eqref{eq:newton-iteration}. By Proposition~\ref{prop:stages}, its cost is two, with one query of order zero and another of order one, both at $x$.

\par\medskip
\needspace{8\baselineskip}

\par\medskip
\needspace{8\baselineskip}

\section{Simulation of Queries under Changes of Variables}\label{sec:simulacion}

Consider procedures querying a composite function, although we have access only to queries to the original function. We shall determine the data indispensable for reproducing their responses exactly, even when other nodes or orders are selected adaptively.

\subsection{Function Class and Known Change of Variables}\label{subsec:chi-modelo}

Let $J,K\subseteq\R$ be nonempty open intervals. Fix a diffeomorphism of class $C^\infty$ \[
\chi:J\longrightarrow K.
\] This means that $\chi$ is bijective and both $\chi$ and $\chi^{-1}$ are of class $C^\infty$. Differentiating $\chi^{-1}(\chi(t))=t$, we obtain \[
(\chi^{-1})'(\chi(t))\chi'(t)=1,
\qquad t\in J.
\] Therefore, \begin{equation}\label{eq:chi-nozero}
\chi'(t)\ne0\qquad(t\in J).
\end{equation}

We denote by $\mathcal F_K$ the class of functions supplied together with their domains \[
f:I_f\longrightarrow\R,
\qquad \varnothing\ne I_f\subseteq K,
\qquad I_f\text{ open interval},
\qquad f\in C^\infty(I_f).
\] The domain is part of the object $f$, but is not supplied to the procedure. In particular, the same expression restricted to two different intervals represents two distinct inputs. For each $f\in\mathcal F_K$, define \begin{equation}\label{eq:chi-gdomain}
g_f:=f\circ\chi,
\qquad I_{g_f}:=\chi^{-1}(I_f).
\end{equation} Since $\chi$ is a homeomorphism between intervals, $I_{g_f}$ is a nonempty open interval. Moreover, $g_f\in C^\infty(I_{g_f})$ and \begin{equation}\label{eq:chi-domain-test}
t\in I_{g_f}
\quad\Longleftrightarrow\quad
t\in J\text{ and }\chi(t)\in I_f.
\end{equation}

The function $\chi$ and its derivatives are known, and operations involving them are internal. Responses from $f$ are acquired exclusively through the scalar queries of Definition~\ref{def:query}. We do not require computability of the rules, in particular of selecting coefficients that are exactly nonzero; nor is the interval $I_f$ known in advance.

\subsection{Composition Coefficients and Active Data}\label{subsec:chi-coeficientes}

We define functions $b_{r,k}:J\to\R$, for $r\in\Nn$ and $k\in\mathbb Z$, as follows: \[
b_{0,0}=1,
\qquad b_{0,k}=0\quad(k\ne0),
\] and successively \begin{equation}\label{eq:chi-recursion}
b_{r+1,k}=b'_{r,k}+\chi' b_{r,k-1},
\qquad 0\le k\le r+1.
\end{equation} For each $r$, we take $b_{r,k}=0$ when $k<0$ or $k>r$. By induction, all the coefficients are of class $C^\infty$ on $J$.

\begin{lemma}[Composition formula]\label{lem:chi-chain}
For every $f\in\mathcal F_K$, $t\in I_{g_f}$, and $r\in\Nn$, we have \begin{equation}\label{eq:chi-chain}
g_f^{(r)}(t)=\sum_{k=0}^{r}b_{r,k}(t)f^{(k)}(\chi(t)).
\end{equation} Moreover, for $r\ge1$, \begin{equation}\label{eq:chi-diagonal}
b_{r,0}=0,
\qquad b_{r,r}=(\chi')^r.
\end{equation}
\end{lemma}

\begin{proof}
\proofstep{1. Proof of the formula.}
For $r=0$, the equality is the definition $g_f(t)=f(\chi(t))$. Suppose it holds for some $r\in\Nn$. Since all the functions involved are smooth in a neighborhood of $t$, we may differentiate the finite sum and apply the chain rule. We obtain \[
\begin{aligned}
g_f^{(r+1)}(t)
&=\sum_{k=0}^{r}b'_{r,k}(t)f^{(k)}(\chi(t))
 +\sum_{k=0}^{r}b_{r,k}(t)\chi'(t)f^{(k+1)}(\chi(t))\\
&=\sum_{k=0}^{r+1}
\bigl(b'_{r,k}(t)+\chi'(t)b_{r,k-1}(t)\bigr)
 f^{(k)}(\chi(t))\\
&=\sum_{k=0}^{r+1}b_{r+1,k}(t)f^{(k)}(\chi(t)).
\end{aligned}
\] In the second equality, we used the convention of zero coefficients outside their index ranges. Thus, induction proves~\eqref{eq:chi-chain}.

\proofstep{2. Extreme coefficients.}
For $r=1$, the recurrence gives $b_{1,0}=0$ and $b_{1,1}=\chi'$. If $b_{r,0}=0$, then \[
b_{r+1,0}=b'_{r,0}+\chi'b_{r,-1}=0.
\] On the other hand, since $b_{r,r+1}=0$, we have \[
b_{r+1,r+1}=\chi'b_{r,r}.
\] A second induction gives $b_{r,r}=(\chi')^r$. We conclude the two identities in~\eqref{eq:chi-diagonal}.
\end{proof}

The preceding formula is a recursive form of the classical Fa\`a di Bruno formula; see, for example,~\cite[Section 2]{scholzberger}. We do not present the composition identity or its coefficients as new results. The question we shall study is the necessity of acquiring the data appearing with nonzero coefficients.

\begin{definition}[Active orders and queries]\label{def:chi-active}
For $t\in J$ and $r\in\Nn$, we define \[
A_\chi(t,r):=\{k\in\{0,\ldots,r\}:b_{r,k}(t)\ne0\}
\] and \[
\mathcal B_\chi(t,r):=
\{(\chi(t),k):k\in A_\chi(t,r)\}.
\] The elements of $A_\chi(t,r)$ are called active orders of the query $(t,r)$; the elements of $\mathcal B_\chi(t,r)$ are the corresponding active queries to $f$.
\end{definition}

In particular, \[
A_\chi(t,0)=\{0\},
\qquad r\in A_\chi(t,r)\quad(r\ge1),
\] since $\chi'(t)\ne0$. Therefore, all these sets are finite and nonempty. Note that the active set may depend on the point $t$: we shall not replace $A_\chi(t,r)$ by the set of coefficients that are not identically zero throughout $J$.

\subsection{Perturbations Preserving Acquired Responses}\label{subsec:chi-hermite}

Before proving the lower bound, we establish the two properties that we shall use to preserve a computation.

\begin{lemma}[Prescription of finite derivative data]\label{lem:chi-hermite}
Let $a_1,\ldots,a_\ell$ be distinct real points and $m_1,\ldots,m_\ell\in\Nn$. Define \[
M:=\sum_{i=1}^{\ell}(m_i+1).
\] Given real numbers $c_{i,k}$, with $1\le i\le\ell$ and $0\le k\le m_i$, there exists a unique polynomial $h$ of degree less than $M$ such that \[
h^{(k)}(a_i)=c_{i,k}.
\] Consequently, derivative values at any finite set of distinct pairs $(a,k)$ can be prescribed freely.
\end{lemma}

\begin{proof}
\proofstep{1. Injectivity of the data mapping.}
Denote by $\PP_{M-1}$ the space of real polynomials of degree at most $M-1$, including the zero polynomial. Consider the linear mapping \[
\mathcal H:\PP_{M-1}\longrightarrow\R^M,
\qquad
\mathcal H(h)=\bigl(h^{(k)}(a_i)\bigr)_{i,k}.
\] Suppose that $\mathcal H(h)=0$. For each $i$, the derivatives of $h$ at $a_i$ vanish up to order $m_i$. Therefore, $(u-a_i)^{m_i+1}$ divides $h(u)$. Since the points are distinct, the corresponding factors are pairwise coprime, so \[
\prod_{i=1}^{\ell}(u-a_i)^{m_i+1}
\quad\text{divides }h(u).
\] The product has degree $M$. If $h$ were nonzero, its degree would be at least $M$, contradicting $h\in\PP_{M-1}$. Thus, $h=0$ and $\mathcal H$ is injective.

\proofstep{2. Existence and uniqueness.}
The spaces $\PP_{M-1}$ and $\R^M$ have the same dimension $M$. An injective linear mapping between them is surjective. Therefore, every vector $(c_{i,k})_{i,k}$ has a unique preimage. This proves the main assertion.

\proofstep{3. Nonconsecutive orders.}
Given a nonempty finite set of distinct pairs, group the pairs by their nodes. At each node, take $m_i$ to be the largest prescribed order and assign, for example, the value zero to intermediate orders not prescribed. Applying the preceding steps, we obtain a polynomial with all the requested data. If the set of prescriptions is empty, we may take $h=0$. We conclude the final assertion.
\end{proof}

This lemma is the classical Hermite interpolation property. Its proof is included to make precise the freedom of perturbations, not as an independent contribution.

\begin{lemma}[Preservation of a valid computation]\label{lem:chi-keep-run}
Let $\mathscr S$ be a computational procedure and let $f,\widetilde f\in C^\infty(I)$ be supplied on the same open interval $I$. Suppose that $\mathscr S$ has a valid computation on $(f,x)$ and that \[
f^{(k)}(z)=\widetilde f^{(k)}(z)
\] for every pair $(z,k)$ queried in that computation. Then the procedure has, on $(\widetilde f,x)$, exactly the same sequence of states, the same queries and responses, and the same output.
\end{lemma}

\begin{proof}
Let $(s_0,\ldots,s_L)$ be the valid computation on $(f,x)$. Initialization depends only on $x$, so the initial state is the same for both functions. Suppose that the states coincide up to an index $i<L$.

If $s_i$ is internal, the next state is $U(s_i)$, which does not depend on the function supplied. If $s_i$ is a query state, the pair $Q(s_i)=(z,k)$ is the same in both executions. Since the original computation is valid, $z\in I$. The query is also available for $\widetilde f$ and, by hypothesis, its response coincides. Therefore, membership in $D_V$ and the state $V(s_i,f^{(k)}(z))$ are the same.

Thus, by induction, the entire original sequence is a valid computation on $(\widetilde f,x)$. Lemma~\ref{lem:calcunique} guarantees its uniqueness. The final state and output coincide. Moreover, any fixed mapping of the final state has the same value in both computations.
\end{proof}

\subsection{Exact Cost of a Transformed Query}\label{subsec:chi-single}

\begin{definition}[Implementation of a transformed query]\label{def:chi-single}
Fix $r\in\Nn$. An exact implementation of order $r$ under $\chi$ is a procedure $\mathscr A$ querying only $f$ and satisfying, for every $f\in\mathcal F_K$, \[
\Omega_{\mathscr A,f}=I_{g_f},
\qquad
A_f(t)=g_f^{(r)}(t)\quad(t\in I_{g_f}).
\] The initial point of this procedure is $t$. Its rules may depend on $\chi$ and $r$, but not on $f$ or $I_f$.
\end{definition}

\begin{theorem}[Exact cost of a transformed response]\label{thm:chi-single}
For each $r\in\Nn$ there exists an exact implementation $\mathscr A_{\chi,r}$ that, on every valid input $(f,t)$, queries each pair in $\mathcal B_\chi(t,r)$ exactly once and makes no other queries.

Conversely, every exact implementation $\mathscr A$ must query all pairs in $\mathcal B_\chi(t,r)$ on every valid input. In particular, \[
N_{\mathscr A}(f,t)\ge |A_\chi(t,r)|,
\qquad
N_{\mathscr A_{\chi,r}}(f,t)=|A_\chi(t,r)|.
\] These conclusions allow adaptive queries of arbitrary orders at other nodes.
\end{theorem}

\begin{proof}
\proofstep{1. Construction and availability.}
If $t\notin J$, we make the procedure block without queries. If $t\in J$, write \[
A_\chi(t,r)=\{k_1<\cdots<k_p\},
\qquad p\ge1.
\] The procedure successively acquires \[
f^{(k_1)}(\chi(t)),\ldots,f^{(k_p)}(\chi(t))
\] and produces the sum \[
\sum_{i=1}^{p}b_{r,k_i}(t)f^{(k_i)}(\chi(t)).
\] The rules are fixed by $\chi,r$ and the initial datum $t$. A formalization by states consists in storing $(t,i,v)$, where $i$ identifies the next query and $v$ the accumulated sum. Incorporating response $\eta$ replaces $v$ by $v+b_{r,k_i}(t)\eta$ and increases $i$ by one; after the last datum, an output state with value $v$ is reached. All these are mappings permitted by Definition~\ref{def:calc}.

If $t\in I_{g_f}$, all data are available and the output is $g_f^{(r)}(t)$ by Lemma~\ref{lem:chi-chain}. If $t\in J\setminus I_{g_f}$, then $\chi(t)\notin I_f$; since $p\ge1$, the first query receives no response and there is no valid computation. By~\eqref{eq:chi-domain-test}, the domain of the procedure is exactly $I_{g_f}$. The number of queries on a valid input is $p=|A_\chi(t,r)|$.

\proofstep{2. Assumption that an active datum is omitted.}
Let $\mathscr A$ be another exact implementation. Fix $f\in\mathcal F_K$ and $t\in I_{g_f}$. Its computation is finite. Denote by $E$ the set of distinct pairs queried in that computation. Suppose there exists $k_*\in A_\chi(t,r)$ such that \[
q_*:=(\chi(t),k_*)\notin E.
\] On the finite set $E\cup\mathcal B_\chi(t,r)$, prescribe the derivatives of a polynomial $h$ as follows: \[
h^{(k_*)}(\chi(t))=1,
\qquad
h^{(k)}(z)=0
\quad\bigl((z,k)\in(E\cup\mathcal B_\chi(t,r))\setminus\{q_*\}\bigr).
\] The prescriptions are compatible because $q_*$ does not belong to $E$. Lemma~\ref{lem:chi-hermite} provides the polynomial $h$.

\proofstep{3. Preservation of the computation and change in the required response.}
Define $\widetilde f=f+h|_{I_f}$ on the same interval $I_f$. Then $\widetilde f\in\mathcal F_K$ and $I_{g_{\widetilde f}}=I_{g_f}$. At every pair in $E$, the responses of $f$ and $\widetilde f$ coincide. By Lemma~\ref{lem:chi-keep-run}, $\mathscr A$ produces exactly the same output on $(f,t)$ and $(\widetilde f,t)$.

However, linearity of differentiation and~\eqref{eq:chi-chain} give \[
\begin{aligned}
g_{\widetilde f}^{(r)}(t)-g_f^{(r)}(t)
&=(h\circ\chi)^{(r)}(t)\\
&=\sum_{k=0}^{r}b_{r,k}(t)h^{(k)}(\chi(t))\\
&=b_{r,k_*}(t)\ne0.
\end{aligned}
\] The last inequality holds because $k_*$ is active. For $r=0$ the same computation uses $b_{0,0}=1$. Thus, the same output cannot be correct for both inputs, contradicting the exactness of $\mathscr A$.

\proofstep{4. Conclusion about the cost.}
No active pair can be omitted. The pairs in $\mathcal B_\chi(t,r)$ are distinct, so every implementation makes at least $|A_\chi(t,r)|$ queries. The procedure in Step 1 attains this number on every valid input. This proves the result.
\end{proof}

In view of the theorem, we may write \begin{equation}\label{eq:chi-qcost}
q_\chi(t,r):=|A_\chi(t,r)|
\end{equation} for the pointwise minimum cost, attained by the same implementation for every admissible $f$.

\subsection{Virtual Record and Faithful Simulation}\label{subsec:chi-faithful}

To state the corresponding result for a complete procedure, we must distinguish its two records. A procedure operating on $g_f$ requests pairs $(t,r)$; a simulation with access to $f$ requests pairs $(z,k)$. Their acquisition records need not coincide.

Let $\mathcal L:=\mathcal M$ be the set of finite lists of answered queries defined in the proof of Lemma~\ref{lem:cache}. We identify the element $((t,r),\eta)$ with the triple $(t,r,\eta)$. The empty list belongs to $\mathcal L$.

\begin{definition}[Annotated procedure and faithful simulation]\label{def:chi-faithful}
An annotated procedure is a pair $(\mathscr S,\Lambda)$, where $\mathscr S$ is a procedure from Definition~\ref{def:calc} and \[
\Lambda:\Sigma_{\mathrm{sal}}^{\mathscr S}\longrightarrow\mathcal L
\] is a fixed mapping. No query to $f$ is made to apply $\Lambda$. The procedure still produces a real output; the annotation allows a list to be read from the final state.

Fix a procedure $\mathscr C$. We say that $(\mathscr S,\Lambda)$ is a faithful simulation of $\mathscr C$ under $\chi$ if, for every $f\in\mathcal F_K$, \begin{equation}\label{eq:chi-faithful-domain}
\Omega_{\mathscr S,f}=\Omega_{\mathscr C,g_f},
\qquad
S_f(x)=C_{g_f}(x)\quad(x\in\Omega_{\mathscr C,g_f}),
\end{equation} and, for each valid input and its final state $\sigma_{\mathrm{fin}}$, \begin{equation}\label{eq:chi-faithful-record}
\Lambda(\sigma_{\mathrm{fin}})=\Rec_{\mathscr C,g_f}(x).
\end{equation} We call the list on the right the virtual record and $\Rec_{\mathscr S,f}(x)$ the effective record. The mapping $\Lambda$, like all the rules of $\mathscr S$, is fixed independently of $f$.
\end{definition}

The definition requires reconstructing the entire virtual record in every valid computation. It does not require publishing each response during execution; therefore, it also allows simulations that acquire data in advance. The construction we give satisfies a stronger property: it incorporates each virtual response before continuing the corresponding state of $\mathscr C$.

For an input $x\in\Omega_{\mathscr C,g_f}$, write \[
\Rec_{\mathscr C,g_f}(x)
=\bigl(t_j,r_j,\eta_j\bigr)_{j=1}^{\ell},
\qquad
\eta_j=g_f^{(r_j)}(t_j).
\] We define the set of active data of that execution by \begin{equation}\label{eq:chi-union}
\mathcal U_\chi(\mathscr C;f,x)
:=\bigcup_{j=1}^{\ell}\mathcal B_\chi(t_j,r_j).
\end{equation} If $\ell=0$, take $\mathcal U_\chi(\mathscr C;f,x)=\varnothing$. The virtual list is finite, so this set is also finite, although there may be no uniform bound on its cardinality.

\subsection{Exact Simulation of a Complete Procedure}\label{subsec:chi-full}

\begin{theorem}[Faithful simulation and counting by union of queries]\label{thm:chi-simulation}
Let $\mathscr C$ be a computational procedure. There exists a faithful simulation $(\mathscr S_\chi,\Lambda_\chi)$, constructed from $\mathscr C$ and $\chi$, that makes exactly one acquisition for each pair in $\mathcal U_\chi(\mathscr C;f,x)$ and no others on each valid input. Therefore, \begin{equation}\label{eq:chi-union-cost}
N_{\mathscr S_\chi}(f,x)
=|\mathcal U_\chi(\mathscr C;f,x)|.
\end{equation} Moreover, for every $k\in\Nn$, \begin{equation}\label{eq:chi-profile}
N_{\mathscr S_\chi,k}(f,x)
=\bigl|\{z:(z,k)\in\mathcal U_\chi(\mathscr C;f,x)\}\bigr|.
\end{equation} No uniform budget is required for $\mathscr C$, nor any uniform bound on the orders of its queries.
\end{theorem}

\begin{proof}
\proofstep{1. Construction of a simulation before reusing data.}
We use the notation from Definition~\ref{def:calc} for the components of $\mathscr C$. We first construct a procedure $\mathscr S_\chi^0$. We use distinct tags $\mathtt e$ and $\mathtt w$. An execution state is \[
e(s,H):=(\mathtt e,s,H),\qquad s\in\Sigma,\quad H\in\mathcal L.
\] Here $s$ is the virtual state and $H$ contains the virtual responses already incorporated.

For $s\in\Sigma_{\mathrm{con}}$ with $Q(s)=(t,r)$ and $t\in J$, write \[
A_\chi(t,r)=\{k_1<\cdots<k_p\}.
\] We add the working states \[
w(s,H,i,v):=(\mathtt w,s,H,i,v),
\qquad 1\le i\le p+1,\quad v\in\R.
\] The value $v$ will be the accumulated sum for the virtual response. The state set is the disjoint union of all the described $e$ and $w$ states.

We classify $e(s,H)$ with $s\in\Sigma_{\mathrm{sal}}$ as output states. We classify the following as internal states: \[
\begin{array}{ll}
e(s,H),&s\in\Sigma_{\mathrm{int}},\\
e(s,H),&s\in\Sigma_{\mathrm{con}},\ Q(s)=(t,r),\ t\in J,\\
w(s,H,p+1,v),&(s,v)\in D_V.
\end{array}
\] The query states are the $w(s,H,i,v)$ with $1\le i\le p$. The remaining states are blocking states. In particular, a virtual query state with $t\notin J$ cannot continue, nor can a state $w(s,H,p+1,v)$ with $(s,v)\notin D_V$.

These classes are disjoint: the tags separate $e$ and $w$ states; among the former, the original disjoint classes are used, and among the latter, $i\le p$ and $i=p+1$ are separated.

\proofstep{2. Initialization, transitions, queries, and outputs.}
Take $D_\iota$ as the initialization domain and define \[
\iota_\chi^0(x)=e(\iota(x),()).
\] The internal transitions are \[
\begin{aligned}
e(s,H)&\longmapsto e(U(s),H),&&s\in\Sigma_{\mathrm{int}},\\
e(s,H)&\longmapsto w(s,H,1,0),
&&s\in\Sigma_{\mathrm{con}},\ Q(s)=(t,r),\ t\in J,\\
w(s,H,p+1,v)&\longmapsto
 e\bigl(V(s,v),H\mathbin\Vert(t,r,v)\bigr),
&&(s,v)\in D_V.
\end{aligned}
\] In the last line, $\Vert$ means appending a triple to the end of the list.

At a query state, define \[
Q_\chi^0(w(s,H,i,v))=(\chi(t),k_i).
\] Incorporation is defined for every real response $\eta$ and satisfies \[
V_\chi^0(w(s,H,i,v),\eta)
=w\bigl(s,H,i+1,v+b_{r,k_i}(t)\eta\bigr).
\] Finally, \[
A_\chi^0(e(s,H))=A(s),
\qquad \Lambda_\chi^0(e(s,H))=H
\quad(s\in\Sigma_{\mathrm{sal}}).
\] All sets and mappings are fixed from $\mathscr C$ and $\chi$. They do not use $f$, $I_f$, or a root. Therefore, we have defined a procedure and an annotation in the sense of the preceding definitions.

\proofstep{3. Preservation of the original valid computations.}
Fix $f\in\mathcal F_K$ and $x\in\Omega_{\mathscr C,g_f}$. Follow the valid computation of $\mathscr C$. Each internal transition is reproduced by the first rule of Step 2. At a query state $s$, validity implies \[
Q(s)=(t,r),\qquad t\in I_{g_f}.
\] By~\eqref{eq:chi-domain-test}, $t\in J$ and $\chi(t)\in I_f$. Thus, the $p$ queries in the working block are available. Once $i-1$ responses have been incorporated, induction on $i$ gives \[
v=\sum_{a=1}^{i-1}b_{r,k_a}(t)f^{(k_a)}(\chi(t)).
\] At $i=p+1$, Lemma~\ref{lem:chi-chain} yields $v=g_f^{(r)}(t)$. Since the original computation is valid, $(s,v)\in D_V$. The last transition in the block appends exactly $(t,r,g_f^{(r)}(t))$ to the record and produces the correct next virtual state.

Each block has finitely many states, since $p\le r+1<\infty$. The original computation is also finite. Therefore, we reach an output state $e(s_{\mathrm{fin}},H)$ with \[
A_\chi^0(e(s_{\mathrm{fin}},H))=C_{g_f}(x),
\qquad H=\Rec_{\mathscr C,g_f}(x).
\] This proves one inclusion of domains, together with the corresponding equalities.

\proofstep{4. No additional valid computations are introduced.}
Now suppose that $\mathscr S_\chi^0$ has a valid computation on $(f,x)$. Its initial state is $e(\iota(x),())$, and an output state has the form $e(s,H)$ with $s\in\Sigma_{\mathrm{sal}}$.

Between two consecutive states of type $e$, the procedure reproduces an original internal transition or completes a block $w$. In the latter case, all queries in the block received responses. Since $p\ge1$, this implies $\chi(t)\in I_f$; moreover, entry into the block already required $t\in J$. Therefore, $t\in I_{g_f}$. The accumulated value is again $g_f^{(r)}(t)$. Exiting the block requires $(s,g_f^{(r)}(t))\in D_V$ and applies the correct original transition.

Compressing these finite blocks yields a finite sequence of states satisfying Definition~\ref{def:validcalc} for $\mathscr C$ on $(g_f,x)$ and ending in $\Sigma_{\mathrm{sal}}$. Thus, $\Omega_{\mathscr S_\chi^0,f}=\Omega_{\mathscr C,g_f}.$ The annotation and output coincide with the originals. This proof includes procedures that may block or have infinite executions: if the simulation reached a finite output, compression would produce a finite valid original computation.

\proofstep{5. Reuse of effective responses.}
Apply Lemma~\ref{lem:cache} to procedure $\mathscr S_\chi^0$. Denote the resulting procedure by $\mathscr S_\chi$. The lemma preserves its domain and output and makes only one acquisition for each distinct pair originally queried.

In the construction of that lemma, the states are pairs $(\sigma,M)$, where $\sigma$ is a state of the procedure being simulated and $M$ contains the stored responses. At output states, define \[
\Lambda_\chi(\sigma,M):=\Lambda_\chi^0(\sigma).
\] By Step 4 of the proof of Lemma~\ref{lem:cache}, the first component of the final state is preserved; hence so is the annotation. The memory is initialized empty in each computation.

\proofstep{6. Total count and count by orders.}
At each virtual query $(t_j,r_j)$, the procedure before reuse requests exactly the pairs in $\mathcal B_\chi(t_j,r_j)$. The set of distinct pairs in its entire computation is, by definition, $\mathcal U_\chi(\mathscr C;f,x)$. Lemma~\ref{lem:cache} then gives~\eqref{eq:chi-union-cost} and~\eqref{eq:chi-profile}.

If the virtual computation makes no queries, there are no blocks and both counts are zero. This completes the proof.
\end{proof}

\subsection{Optimality among All Faithful Simulations}\label{subsec:chi-optimalidad}

\begin{theorem}[Necessity of every active datum in the record]\label{thm:chi-optimal}
Let $(\mathscr S,\Lambda)$ be any faithful simulation of $\mathscr C$ under $\chi$. For every $f\in\mathcal F_K$ and $x\in\Omega_{\mathscr C,g_f}$, the computation of $\mathscr S$ on $(f,x)$ must query all pairs in $\mathcal U_\chi(\mathscr C;f,x)$.

Consequently, the simulation of Theorem~\ref{thm:chi-simulation} is optimal at every valid input, simultaneously for the total number and the number of acquisitions of each order, among all faithful simulations.
\end{theorem}

\begin{proof}
\proofstep{1. Effective record and assumption of omission.}
Fix a valid input $(f,x)$ and write \[
\Rec_{\mathscr C,g_f}(x)
=\bigl(t_j,r_j,\eta_j\bigr)_{j=1}^{\ell}.
\] Denote by $E$ the finite set of distinct pairs queried by $\mathscr S$ on $(f,x)$. If $\ell=0$, the active set is empty and the asserted inclusion is immediate. Suppose there exists \[
q_*=(a,k_*)\in\mathcal U_\chi(\mathscr C;f,x)\setminus E.
\] By Lemma~\ref{lem:chi-hermite}, there exists a polynomial $h$ whose derivatives satisfy \[
h^{(k_*)}(a)=1,
\qquad
h^{(k)}(z)=0
\quad\bigl((z,k)\in(E\cup\mathcal U_\chi(\mathscr C;f,x))\setminus\{q_*\}\bigr).
\] The prescriptions are compatible, since $q_*$ does not appear in $E$.

\proofstep{2. The simulator cannot distinguish the perturbation.}
Define $\widetilde f=f+h|_{I_f}$. The functions $f$ and $\widetilde f$ are supplied on the same interval and both belong to $\mathcal F_K$. Lemma~\ref{lem:chi-keep-run} applied to $\mathscr S$ shows that its computation on $(\widetilde f,x)$ has the same sequence of states as on $(f,x)$. In particular, it terminates in the same state $\sigma_{\mathrm{fin}}$ and has the same annotation $\Lambda(\sigma_{\mathrm{fin}})$.

By fidelity, both original computations are valid and their records should satisfy \begin{equation}\label{eq:chi-same-virtual}
\Rec_{\mathscr C,g_{\widetilde f}}(x)
=\Lambda(\sigma_{\mathrm{fin}})
=\Rec_{\mathscr C,g_f}(x).
\end{equation}

\proofstep{3. First virtual response affected.}
Since $q_*$ belongs to the active union, there is a least index $j_*$ such that \[
q_*\in\mathcal B_\chi(t_{j_*},r_{j_*}).
\] For $j<j_*$, all derivatives of $h$ actively entering query $(t_j,r_j)$ vanish. By~\eqref{eq:chi-chain}, \[
g_{\widetilde f}^{(r_j)}(t_j)=g_f^{(r_j)}(t_j)
\qquad(j<j_*).
\] By induction on that prefix of the original computation, internal transitions and query decisions remain the same until the query at index $j_*$ is reached. We use no continuity of the rules; we use exact equality of preceding responses.

At that query, $a=\chi(t_{j_*})$ and $k_*$ is active. The prescriptions for $h$ give \[
\begin{aligned}
g_{\widetilde f}^{(r_{j_*})}(t_{j_*})
-g_f^{(r_{j_*})}(t_{j_*})
&=\sum_{k=0}^{r_{j_*}}b_{r_{j_*},k}(t_{j_*})h^{(k)}(\chi(t_{j_*}))\\
&=b_{r_{j_*},k_*}(t_{j_*})\ne0.
\end{aligned}
\] The intervals of $g_f$ and $g_{\widetilde f}$ coincide, so the query remains available. Its response is different. This contradicts the equality of records~\eqref{eq:chi-same-virtual}.

\proofstep{4. Bounds and attainment of the minimum.}
We conclude that $\mathcal U_\chi(\mathscr C;f,x)\subseteq E.$ Therefore, \[
N_{\mathscr S}(f,x)\ge|\mathcal U_\chi(\mathscr C;f,x)|,
\] and, for each $k\in\Nn$, \[
N_{\mathscr S,k}(f,x)
\ge\bigl|\{z:(z,k)\in\mathcal U_\chi(\mathscr C;f,x)\}\bigr|.
\] Theorem~\ref{thm:chi-simulation} provides one simulation attaining all these bounds. Thus, the minima are attained simultaneously at every valid input.
\end{proof}

\begin{remark}[Complete record and final output]\label{rem:chi-not-output}
The comparison class in Theorem~\ref{thm:chi-optimal} requires reproducing the entire virtual record. It is not the class of all procedures that produce only the same final output.

To see the difference, take $\chi(t)=e^t$, fix $r\ge2$, and consider a procedure that requests $g_f^{(r)}(0)$ and then returns zero, without using the value received. Its valid domain is all of $\R$ or the empty set according as $1\in I_f$ or $1\notin I_f$, respectively. A procedure querying only $f(1)$ and returning zero upon receiving a response preserves this domain and output. However, it does not in general reconstruct virtual response $g_f^{(r)}(0)$. In the exponential example of Subsection~\ref{subsec:chi-ejemplos}, we shall see that a faithful simulation requires $r$ acquisitions on every valid input.
\end{remark}

\subsection{Transfer of Budgets to the Query Model}\label{subsec:chi-presupuesto}

\begin{corollary}[Effective budget and exact representation]\label{cor:chi-budget}
Let $d\in\N$ and $s\in\Nn$. Suppose that, for every $f\in\mathcal F_K$ and $x\in\Omega_{\mathscr C,g_f}$, the valid computation of $\mathscr C$ uses at most $d$ queries, all of orders at most $s$. Define \[
B:=d\max\{1,s\}.
\] The simulation of Theorem~\ref{thm:chi-simulation} satisfies \[
N_{\mathscr S_\chi}(f,x)\le B
\] on these inputs. Moreover, there exists a method $T_{\mathscr S_\chi,B}$ with budget $B$ such that, for every $f\in\mathcal F_K$, \[
\Omega_{T_{\mathscr S_\chi,B},f}=\Omega_{\mathscr C,g_f},
\qquad
(T_{\mathscr S_\chi,B})_f(x)=C_{g_f}(x).
\] The method preserves the effective record and counts of $\mathscr S_\chi$. Its history and the fixed rules also allow reconstruction of the virtual record.
\end{corollary}

\begin{proof}
If $r=0$, then $|A_\chi(t,r)|=1$. If $1\le r\le s$, by~\eqref{eq:chi-diagonal}, $A_\chi(t,r)\subseteq\{1,\ldots,r\}$ and \[
|A_\chi(t,r)|\le r\le s.
\] Thus, for a valid list of length $\ell\le d$, \[
\begin{aligned}
N_{\mathscr S_\chi}(f,x)
&\le\sum_{j=1}^{\ell}|A_\chi(t_j,r_j)|\\
&\le d\max\{1,s\}=B.
\end{aligned}
\] Since $B\ge1$, we may apply Theorem~\ref{thm:exactrepresentation} to $\mathscr S_\chi$ and $B$. Its domain equality gives \[
\Omega_{T_{\mathscr S_\chi,B},f}
=\{x\in\Omega_{\mathscr S_\chi,f}:N_{\mathscr S_\chi}(f,x)\le B\}.
\] On the class $\mathcal F_K$, the preceding bound and faithfulness identify this set with $\Omega_{\mathscr C,g_f}$. The same theorem preserves outputs, effective records, and counts.

Finally, if the method terminates after $k$ queries with history $\hc_k=(x,\eta_1,\ldots,\eta_k)$, the construction in the proof of Theorem~\ref{thm:exactrepresentation} gives $b_k(\hc_k)=\sigma_{\mathrm{fin}}$, the final state of $\mathscr S_\chi$. Applying $\Lambda_\chi$ to this state recovers the virtual record without new queries to $f$.
\end{proof}

The virtual budget $d$ and the effective budget $B$ correspond to different oracles and must not be identified.

\subsection{Characterization of Unit-Cost Changes}\label{subsec:chi-affine}

\begin{corollary}[Affine characterization]\label{cor:chi-affine}
For the diffeomorphism $\chi:J\to K$, the following statements are equivalent:
\begin{enumerate}[label=\textup{(\roman*)}]
\item For every $t\in J$ and $r\in\Nn$, $q_\chi(t,r)=1$.
\item For every $t\in J$, $q_\chi(t,2)=1$.
\item There exist $a,b\in\R$, with $a\ne0$, such that $\chi(t)=at+b$ for every $t\in J$.
\end{enumerate}
\end{corollary}

\begin{proof}
\proofstep{$\textup{(iii)}\Longrightarrow\textup{(i)}$.}
Suppose that $\chi(t)=at+b$, with $a\ne0$. By induction using the chain rule, \[
(f\circ\chi)^{(r)}(t)=a^r f^{(r)}(at+b)
\quad(r\ge1).
\] For $r=0$, the response is $f(at+b)$. In both cases, only one datum appears with a nonzero coefficient. Theorem~\ref{thm:chi-single} gives $q_\chi(t,r)=1$.

\proofstep{$\textup{(i)}\Longrightarrow\textup{(ii)}$.}
It suffices to take $r=2$.

\proofstep{$\textup{(ii)}\Longrightarrow\textup{(iii)}$.}
Recurrence~\eqref{eq:chi-recursion} gives \[
b_{2,1}(t)=\chi''(t),
\qquad b_{2,2}(t)=(\chi'(t))^2.
\] The second coefficient never vanishes. If there were $t_0\in J$ with $\chi''(t_0)\ne0$, we would have two active orders and, by Theorem~\ref{thm:chi-single}, $q_\chi(t_0,2)=2$, contradicting \textup{(ii)}. Therefore, $\chi''=0$ on $J$.

Fix $t_0\in J$. Applying the mean value theorem to $\chi'$ on the segment joining $t_0$ to any $t\in J$, we conclude that $\chi'(t)=\chi'(t_0)=:a$. By~\eqref{eq:chi-nozero}, $a\ne0$. The function $\chi(t)-at$ has zero derivative and, applying the mean value theorem again, is constant on the interval. Thus, $\chi(t)=at+b$ throughout $J$. This proves the last implication.
\end{proof}

The characterization concerns all individual queries at all points. It does not assert that every particular procedure has a higher cost under every nonaffine change. A single procedure may reuse data between distinct queries, as we shall see below.

\subsection{Examples: Individual Queries and Complete Blocks}\label{subsec:chi-ejemplos}

\begin{example}[Exponential change]\label{ex:chi-exp}
Consider $\chi(t)=e^t$, from $\R$ to $(0,\infty)$. Define numbers $S(r,k)$ by \[
S(0,0)=1,
\qquad
S(r+1,k)=kS(r,k)+S(r,k-1),
\] with value zero outside indices $0\le k\le r$ of each row. For $r\ge1$, we have $S(r,0)=0$, $S(r,1)=S(r,r)=1$, and $S(r,k)>0$ when $1\le k\le r$. Indeed, the equalities at the endpoints follow directly from the recurrence, and the interior terms are sums of nonnegative terms with at least one positive term; induction on $r$ proves positivity.

We show that \begin{equation}\label{eq:chi-stirling}
b_{r,k}(t)=S(r,k)e^{kt}.
\end{equation} The equality holds for $r=0$. If it holds in one row, then \[
\begin{aligned}
b_{r+1,k}(t)
&=kS(r,k)e^{kt}+e^tS(r,k-1)e^{(k-1)t}\\
&=\bigl(kS(r,k)+S(r,k-1)\bigr)e^{kt}\\
&=S(r+1,k)e^{kt}.
\end{aligned}
\] This proves~\eqref{eq:chi-stirling}. Consequently, for $r\ge1$, \[
A_\chi(t,r)=\{1,\ldots,r\},
\qquad q_\chi(t,r)=r.
\] For $r=0$, the cost remains one. In particular, \[
(f\circ\exp)'''(0)=f'(1)+3f''(1)+f'''(1),
\] and every exact implementation must acquire these three data. It cannot replace them with two queries of other orders or at other nodes.
\end{example}

\begin{example}[Quadratic change on the positive half-line]\label{ex:chi-square}
Let $\chi(t)=t^2$ be from $(0,\infty)$ to $(0,\infty)$. For $r\ge1$, \[
b_{r,k}(t)=
\frac{r!}{(2k-r)!(r-k)!}(2t)^{2k-r}
\quad\left(\left\lceil\frac r2\right\rceil\le k\le r\right),
\] and the remaining coefficients are zero.

To verify the formula, fix $t>0$ and write $\chi(t+h)=t^2+2th+h^2$. The Taylor expansion of $f$ up to order $r$ at $t^2$ gives \[
f(t^2+2th+h^2)
=\sum_{k=0}^{r}\frac{f^{(k)}(t^2)}{k!}(2th+h^2)^k+o(|h|^r).
\] Indeed, $2th+h^2=O(h)$ and the composed remainder is $o(|h|^r)$. In $(2th+h^2)^k$, a term of degree $r$ is obtained by choosing $r-k$ factors $h^2$ and $2k-r$ factors $2th$. This is possible exactly when $\lceil r/2\rceil\le k\le r$, and its coefficient is \[
\binom{k}{r-k}(2t)^{2k-r}.
\] To justify the identification, apply Lemma~\ref{lem:chi-chain} to $F_k(y)=(y-t^2)^k/k!$. Its only nonzero derivative at $t^2$ is of order $k$, equal to one. Thus $b_{r,k}(t)$ is $r!$ times the coefficient of $h^r$ in $(2th+h^2)^k/k!$, giving the asserted formula. Since $t>0$, all the indicated coefficients are nonzero. Therefore, \[
A_\chi(t,r)=\{\lceil r/2\rceil,\ldots,r\},
\qquad q_\chi(t,r)=\lfloor r/2\rfloor+1.
\] This count is exact in the scalar-query model, by Theorem~\ref{thm:chi-single}.
\end{example}

\begin{corollary}[Reuse for a complete block]\label{cor:chi-full-block}
Fix $m\in\Nn$ and a node $t\in J$. To reproduce the responses of all queries of orders $0,\ldots,m$ to $g_f$ at $t$, the minimum number of acquisitions from $f$ is $m+1$ on every valid input. The required data are exactly \[
f(\chi(t)),f'(\chi(t)),\ldots,f^{(m)}(\chi(t)).
\] The statement concerns reproduction of the block of responses, not only a final combination of them. Here reproduction means that an annotated procedure, as in Definition~\ref{def:chi-faithful}, has domain $I_{g_f}$ and recovers the ordered list $(t,r,g_f^{(r)}(t))_{r=0}^m$ from its final state.
\end{corollary}

\begin{proof}
For every $r\le m$, we have $A_\chi(t,r)\subseteq\{0,\ldots,r\}$. Moreover, $r\in A_\chi(t,r)$, including $r=0$. Therefore, \[
\bigcup_{r=0}^{m}A_\chi(t,r)=\{0,\ldots,m\}.
\] We may consider the virtual procedure requesting these $m+1$ data, storing their responses in its state, and terminating. Theorems~\ref{thm:chi-simulation} and~\ref{thm:chi-optimal} show that the union of its active queries is necessary and sufficient. Its cardinality is $m+1$. Thus, all responses are obtained, with one acquisition for each indicated datum.
\end{proof}

In the exponential case, answering the orders $0,\ldots,m$ separately, each time discarding the previous information, would use \[
1+\sum_{r=1}^{m}r=1+\frac{m(m+1)}2
\] queries. The procedure reusing data requires only $m+1$. On the other hand, if only orders $0$ and $m\ge1$ are requested, the two virtual queries already require $m+1$ effective data under that same change. These examples show why the cost of a record must be computed from a union of sets rather than by simply adding individual costs.

\begin{corollary}[The virtual budget alone does not control effective cost]\label{cor:chi-unbounded}
For the fixed change $\chi(t)=e^t$, no uniform bound on the cost of faithful simulation can depend only on the number of virtual queries and hold for all procedures without restricting their derivative orders.
\end{corollary}

\begin{proof}
For each $r\in\N$, consider the procedure that requests $g_f^{(r)}(0)$ and returns that response. Its virtual budget is one. On every input with $1\in I_f$, Example~\ref{ex:chi-exp} and Theorem~\ref{thm:chi-optimal} show that every faithful simulation requires at least $r$ queries to $f$. Since $r$ is arbitrary, no common finite bound exists for all these procedures based only on budget one. This does not involve increasing the number of virtual queries, but varying their order.
\end{proof}

\subsection{Scope of the Simulation Results}\label{subsec:chi-limits}

The optimality proved uses the full class $\mathcal F_K$. In the lower-bound proof, $f$ is replaced by $f+h$ on the same interval. We do not claim that this perturbation preserves a distinguished root, monotonicity, convexity, derivative bounds, or membership in a restricted parametric family. Therefore, the lower bounds do not automatically extend to classes with these additional promises.

The diffeomorphism hypothesis guarantees that each active support is nonempty. To appreciate its role, consider, outside that hypothesis, $\chi(t)=t^3$ on $\R$. When $0\in I_f$, we have $(f\circ\chi)'(0)=0$, although no active datum appears in the first-order formula at that point. A procedure returning zero without queries does not preserve the availability of this response over all unknown intervals: it would also return zero when $0\notin I_f$. A query to $f(0)$, followed by output zero if a response is received, does preserve that domain. Thus, the support formula without a hypothesis guaranteeing active data does not by itself characterize the availability cost.

The cost studied is the number of exact scalar acquisitions. Scholz and Berger~\cite{scholzberger} study the computation of the derivative structure of compositions using Fa\`a di Bruno and Bell polynomials. Naumann~\cite[Sections 1--2]{naumann} considers minimizing arithmetic operations when evaluating chain rules with supplied elementary derivatives. Here these operations are internal and the function $f$ is unknown. These differences of model preclude interpreting our counts as time bounds or as direct comparisons with automatic differentiation algorithms having access to the function's code.

Interpolation, the composition formula, and the information-preservation argument are classical tools. The results established in this section are their specific formulations for faithful simulation of partial procedures, with a supplied domain, a virtual record, reuse, and exact cost.

\par\medskip
\needspace{8\baselineskip}

\section{Composition of Changes of Variables and Modular Cost}\label{sec:composicion-modular}

The faithful simulation in the preceding section preserves the responses requested by a procedure, not merely its final output. This distinction takes a concrete form when successive changes of variables occur: we may reconstruct the responses at each stage or simplify the composition first. Let us see how the costs of these two tasks are related and what happens when a change is undone by its inverse.

\subsection{Precedents and Scope of the Comparison}\label{subsec:comp-antecedentes}

Matrix composition and its inversion were studied by Jabotinsky~\cite[Theorems I--II]{jabotinsky}; see also the Stirling matrices of Aschenbrenner~\cite[Section 3]{aschenbrenner} and the composition chains of Johnston and Prochno~\cite[Theorem 2.1]{johnston}. These identities are classical. Here we count scalar acquisitions from an unknown function, rather than arithmetic operations on available derivatives~\cite{scholzberger,naumann}; the lower bounds use indistinguishability of data~\cite{werschulz} and Lemma~\ref{lem:chi-hermite}.

\subsection{Model, Coefficients, and Reconstruction Tasks}\label{subsec:comp-modelo}

We use the class $\mathcal F_K$ from Subsection~\ref{subsec:chi-modelo} and the scalar queries of Definition~\ref{def:query}. The interval of $f$ remains unknown, and an unavailable query supplies no information. The known changes are processed internally, and the virtual record is recovered through an annotation of the final state, as in Definition~\ref{def:chi-faithful}.

\begin{definition}[Coefficients of a composition]\label{comp:def:coeficientes}
Let $\theta:U\to V$ be a $C^\infty$ diffeomorphism between open intervals. For $n\ge1$ and $1\le k\le n$, we denote by $B^\theta_{n,k}(u)$ the coefficients obtained from construction~\eqref{eq:chi-recursion} by replacing $\chi$ with $\theta$. The composition formula, its recurrence, and the diagonal coefficient are then those of~\eqref{eq:chi-chain} and~\eqref{eq:chi-diagonal}, with the same replacement.
\end{definition}

For $\theta=\chi$, these coefficients coincide with the $b_{n,k}$ of Subsection~\ref{subsec:chi-coeficientes}. The superscript distinguishes the transformations involved in a composition. For a polynomial $P$, we write $[z^q]P$ for its coefficient of degree $q$. The support of a finite family of coefficients is the set of indices where they are nonzero.

The identity follows by induction using the chain rule and is a form of Fa\`a di Bruno's formula. For a $C^\infty$ function, it uses only finitely many derivatives at each order. It does not assume analyticity.

Fix diffeomorphisms \[
 \psi:L\to J,\qquad \chi:J\to K,
\] between open intervals. For $I\subseteq K$, define \[
 g=f\circ\chi,\qquad h=g\circ\psi=f\circ\chi\circ\psi
\] on $\chi^{-1}(I)$ and $(\chi\circ\psi)^{-1}(I)$, respectively. Fix $t\in L$ and $r\in\N$, and write \[
 b=\psi(t),\qquad a=\chi(b),\qquad
 \alpha_{j,k}=B^\chi_{j,k}(b),\qquad \beta_j=B^\psi_{r,j}(t).
\] Then \begin{align}
 g^{(j)}(b)&=\sum_{k=1}^j\alpha_{j,k}f^{(k)}(a),\label{comp:eq:intermedia}\\
 h^{(r)}(t)&=\sum_{j=1}^r\beta_j g^{(j)}(b)
           =\sum_{k=1}^r\gamma_k f^{(k)}(a),\label{comp:eq:final}\\
 \gamma_k&=\sum_{j=k}^r\beta_j\alpha_{j,k}.\label{comp:eq:gamma}
\end{align}

\begin{definition}[Two tasks and three sets of orders]\label{comp:def:tareas}
We define \[
 V=\{j:\beta_j\ne0\},\qquad
 U=\bigcup_{j\in V}\{k:\alpha_{j,k}\ne0\},\qquad
 D=\{k:\gamma_k\ne0\}.
\] The direct task requires the output $h^{(r)}(t)$. The faithful modular task requires the same output and, as a terminal annotation in the sense of Definition~\ref{def:chi-faithful}, the ordered list of triples $(b,j,g^{(j)}(b))$, $j\in V$. It does not require producing the list during execution; advance acquisitions are allowed.

For either task, a procedure receives an arbitrary input $x\in\R$, queries only $f$, and must have, for every $f\in\mathcal F_K$, the domain
\[
\Omega_{\mathscr A,f}=\begin{cases}
\R,&a\in I_f,\\
\varnothing,&a\notin I_f.
\end{cases}
\]
Its rules may depend on the fixed parameters $\chi,\psi,t,r$, but not on $f$ or $I_f$. On each valid input $(f,x)$, the costs $\cd$ and $\cm$ are the minimum numbers of queries among procedures satisfying the respective task for all $f\in\mathcal F_K$. Theorem~\ref{comp:thm:costes} shows that these minima are independent of the valid input. The fixed parameters are omitted from the notation.
\end{definition}

Note that $r\in V\cap U\cap D$, since \[
 \gamma_r=\beta_r\alpha_{r,r}
 =(\psi'(t))^r(\chi'(b))^r\ne0.
\] Thus, all three sets are nonempty and neither task can dispense with every availability check.

\subsection{Indispensable Derivative Data}

The finite prescription of derivatives that we shall use was already proved in Lemma~\ref{lem:chi-hermite}. We do not repeat its proof here; we state the consequence needed for a list of intermediate responses.

\begin{lemma}[Necessity of every active coordinate]\label{comp:lem:necesidad}
Fix $a\in K$ and a finite family of linear combinations \[
 L_j(f)=\sum_{k\in A} c_{j,k}f^{(k)}(a),\qquad 1\le j\le q,
\] where $A\subseteq\Nn$ is finite. Let \[
 A_*=\{k\in A:\text{there exists }j\text{ with }c_{j,k}\ne0\}.
\] Here reconstruction means that the terminal output and a fixed annotation of the final state determine all values $L_j(f)$. Every deterministic procedure reconstructing these values for every smooth function defined on an interval containing $a$ must query each pair $(a,k)$, $k\in A_*$, in every valid execution.
\end{lemma}
\begin{proof}
\proofstep{1. Assumption of omission.}
Fix a valid input and let $E$ be the finite set of pairs queried by the procedure. Suppose that $(a,k_*)\notin E$, with $k_*\in A_*$. Choose $j_*$ such that $c_{j_*,k_*}\ne0$.

\proofstep{2. Perturbation.}
By Lemma~\ref{lem:chi-hermite}, there exists a polynomial $P$ such that \[
 P^{(k_*)}(a)=1,
\] and all its derivatives corresponding to \[
 E\cup\{(a,k):k\in A_*\}\setminus\{(a,k_*)\}
\] are zero. To apply the lemma, we fill in the unprescribed orders up to the maximum at each node with zeros. The conditions are compatible, since the distinguished pair was not queried.

\proofstep{3. Equality of the computation.}
The functions $f$ and $\widetilde f=f+P|_I$ have the same interval of definition and both are admissible. The initial state of the procedure coincides. If the prefix of states and responses coincides, the next decision is the same. An internal transition is fixed; a query in the original record receives the same response by the construction of $P$. By induction, the procedure reaches the same final state, with the same output and the same annotation. This argument allows adaptive nodes and orders and advance acquisitions.

\proofstep{4. Contradiction.}
By contrast, \[
 L_{j_*}(\widetilde f)-L_{j_*}(f)=c_{j_*,k_*}\ne0.
\] The final state cannot correctly provide both values. We conclude that all active pairs are necessary.
\end{proof}

This is a classical information-indistinguishability argument. The annotation is used to discuss several values without changing the real-valued output type of the procedure. No bound on convergence order is involved.

\subsection{Composition Cost and the Exact Savings Criterion}

\begin{theorem}[Pointwise costs and complete cancellation]\label{comp:thm:costes}
With the preceding definitions, \begin{equation}\label{comp:eq:costes}
 \boxed{\cm=|U|,\qquad \cd=|D|,\qquad D\subseteq U.}
\end{equation} The minima are attained simultaneously on all valid inputs. Strict savings are obtained by abandoning modular reconstruction and computing only the final derivative if and only if there exists $k\in U$ such that \[
 \sum_{j=k}^r\beta_j\alpha_{j,k}=0.
\] The number of acquisitions saved is exactly $|U\setminus D|$.
\end{theorem}
\begin{proof}
\proofstep{1. Modular construction.}
Let us query $f^{(k)}(a)$ only once for each $k\in U$. From these responses, we compute all expressions \eqref{comp:eq:intermedia} with $j\in V$ and store their list. We then compute \eqref{comp:eq:final}. This uses $|U|$ acquisitions.

Since $U\ne\varnothing$ and all queries are made at $a$, the procedure receives all responses if and only if $a\in I$. It introduces no additional valid inputs. The intermediate values and list are reconstructed exactly.

\proofstep{2. Modular lower bound.}
Apply Lemma~\ref{comp:lem:necesidad} to the combinations \eqref{comp:eq:intermedia} with $j\in V$. The union of their active coordinates is precisely $U$. Every faithful procedure must acquire its $|U|$ data, even if it completely reorganizes the computation. The preceding step attains this bound.

\proofstep{3. Direct task.}
Let us query only the pairs $(a,k)$ with $k\in D$ and form the final combination. Since $D\ne\varnothing$, availability is again preserved. Lemma~\ref{comp:lem:necesidad}, applied to a single combination, gives the lower bound $|D|$.

\proofstep{4. Comparison.}
If $k\notin U$, then $\beta_j\alpha_{j,k}=0$ for every $j$ and, by \eqref{comp:eq:gamma}, $\gamma_k=0$. Thus, $D\subseteq U$. The difference in cardinalities is $|U\setminus D|$. It is positive exactly when a coordinate active in some intermediate value cancels completely in the final value. This proves all the assertions.
\end{proof}

\begin{remark}
A partial cancellation with $\gamma_k\ne0$ does not save the query of order $k$. The direct and modular tasks require different observable outputs; their cost difference does not contradict the optimality of faithful simulation.
\end{remark}

\subsection{The Exponential and the Logarithm}

\begin{corollary}[Exact separation of one versus $r$ queries]\label{comp:cor:explog}
Let $\chi(u)=e^u$, $\psi(t)=\log t$, and $a>0$. To compute the derivative of order $r\ge1$ of \[
 h=(f\circ\exp)\circ\log=f
\] at $a$, the direct task has cost one, and the modular task of Subsection~\ref{subsec:comp-modelo} has cost $r$: \[
 \cd=1,\qquad \cm=r.
\] Equality of domains is preserved over all open intervals $I\subseteq(0,\infty)$.
\end{corollary}
\begin{proof}
\proofstep{1. Data of the exponential composition.}
Let $b=\log a$ and $g=f\circ\exp$. By~\eqref{eq:chi-stirling} and~\eqref{eq:chi-chain}, the intermediate derivatives are expressed using the Stirling numbers $S(j,k)$ of the second kind, as in Example~\ref{ex:chi-exp}, evaluated at $b=\log a$. In particular, their coefficients are nonzero for all $1\le k\le j$.

\proofstep{2. Intermediate orders required by the logarithm.}
Identity \begin{equation}\label{comp:eq:stirling1}
 h^{(r)}(a)=a^{-r}\sum_{j=1}^r s(r,j)g^{(j)}(\log a)
\end{equation} is obtained inductively using \[
 s(1,1)=1,\qquad s(r+1,j)=s(r,j-1)-r s(r,j).
\] Indeed, differentiating $a^{-r}g^{(j)}(\log a)$ produces $-ra^{-r-1}g^{(j)}$ and $a^{-r-1}g^{(j+1)}$. The numbers $s(r,j)$ are the signed Stirling numbers of the first kind. The recurrence shows that $(-1)^{r-j}s(r,j)>0$ for $1\le j\le r$. Therefore, $V=\{1,\ldots,r\}$.

\proofstep{3. Modular and direct costs.}
Already the response $g^{(r)}(b)$ contains, with nonzero coefficients, all coordinates $f'(a),\ldots,f^{(r)}(a)$. Consequently, \[
 U=\{1,\ldots,r\}.
\] On the other hand, $\exp\circ\log$ is the identity on $(0,\infty)$, so $h^{(r)}(a)=f^{(r)}(a)$ and $D=\{r\}$. Theorem~\ref{comp:thm:costes} gives the stated costs. All acquisitions are made at $a$, and there is at least one, preserving the domain of availability.
\end{proof}

\begin{example}[Order three]\label{comp:ex:orden-tres}
The intermediate responses are \[
 g'(\log a)=af'(a),\qquad
 g''(\log a)=af'(a)+a^2f''(a),
\] \[
 g'''(\log a)=af'(a)+3a^2f''(a)+a^3f'''(a).
\] Their combination is \[
 h'''(a)=a^{-3}\bigl(g'''(\log a)-3g''(\log a)+2g'(\log a)\bigr)=f'''(a).
\] Reconstructing the three intermediate responses requires three acquisitions; computing only the combination requires one. Each intermediate response has a nonzero coefficient in the rule before simplification.
\end{example}

The Stirling identities and matrix inversion are classical~\cite{aschenbrenner}. The corollary assigns them an exact cost in a particular model; it does not turn that algebra into a new identity. If a query supplied the entire block of derivatives up to order $r$ for free, the count would be different: one block would suffice for both tasks.

\subsection{Inverse Changes: A More General Dichotomy}\label{comp:sec:dicotomia}

The preceding separation extends to more general smooth changes through the following lemma.

\begin{lemma}[Growth in the number of active orders]\label{comp:lem:crecimiento}
Let $\chi:J\to K$ be a $C^\infty$ diffeomorphism and fix $b\in J$. If some derivative $\chi^{(m)}(b)$, with $m\ge2$, is nonzero, then \[
 \#\{k\in\{1,\ldots,r\}:B^\chi_{r,k}(b)\ne0\}\longrightarrow+\infty
 \quad(r\to\infty).
\]
\end{lemma}
\begin{proof}
\proofstep{1. First nonlinear derivative.}
Let $m\ge2$ be the least index with $\chi^{(m)}(b)\ne0$. Define $d=m-1$, $c=\chi'(b)\ne0$, and \[
 v_q=\frac{\chi^{(q+1)}(b)}{(q+1)!c}\quad(q\ge1).
\] Then $v_1=\cdots=v_{d-1}=0$ and $v_d\ne0$.

\proofstep{2. Coefficients from a finite Taylor expansion.}
For each $r$, the derivatives up to order $r$ of a composition depend only on the Taylor polynomials up to that order. In particular, \begin{equation}\label{comp:eq:coef-taylor}
 B^\chi_{r,k}(b)=\frac{r!}{k!}[z^r]
 \left(\sum_{q=1}^r\frac{\chi^{(q)}(b)}{q!}z^q\right)^k.
\end{equation} To justify this, apply the coefficient definition to $F(y)=(y-\chi(b))^k/k!$. Its only nonzero derivative at $\chi(b)$ is that of order $k$, equal to one. Therefore, the left-hand side is the derivative of order $r$ of $F\circ\chi$ at $b$. The finite Taylor expansion of $\chi(b+z)-\chi(b)$ gives the right-hand side. No convergence of an infinite series is used.

\proofstep{3. A polynomial for each diagonal.}
Fix $j\ge1$ and put \[
 V_j(z)=\sum_{q=d}^{dj}v_qz^q,
 \qquad
 P_j(X)=\sum_{\ell=0}^j\binom X\ell[z^{dj}]V_j(z)^\ell.
\] Here $\binom X\ell=X(X-1)\cdots(X-\ell+1)/\ell!$ for $\ell\ge1$ and $\binom X0=1$.

Since the lowest degree of $V_j$ is $d$, only powers $\ell\le j$ can contribute. The term $\ell=j$ has \[
 [z^{dj}]V_j(z)^j=v_d^j\ne0.
\] Therefore, $P_j$ is a polynomial of degree exactly $j$, with leading coefficient $v_d^j/j!$.

For $r>dj$, let $k=r-dj\ge1$. Factoring out $cz$ in \eqref{comp:eq:coef-taylor} and using the binomial formula, we obtain \begin{equation}\label{comp:eq:diagonal}
 B^\chi_{r,r-dj}(b)
 =\frac{r!}{(r-dj)!}c^{r-dj}P_j(r-dj).
\end{equation} Coefficients of degrees greater than $dj$ do not affect extraction of the indicated coefficient. The expression is valid for every smooth function, since in each case it uses only a finite Taylor expansion.

\proofstep{4. An arbitrary number of active orders.}
Since $P_j$ is nonzero, it has finitely many real roots. Thus, for every fixed $j$, the expression in \eqref{comp:eq:diagonal} does not vanish when $r$ is sufficiently large. For $j=0$, the diagonal coefficient is $c^r\ne0$.

Let $M\in\N$. Consider $j=0,\ldots,M-1$. There exists $R_M$ such that, for every $r\ge R_M$, the $M$ indices \[
 r,\ r-d,\ r-2d,\ldots,r-d(M-1)
\] are positive and have nonzero coefficients. Consequently, there are at least $M$ active orders for every $r\ge R_M$. Since $M$ was arbitrary, the asserted limit follows.
\end{proof}

\begin{theorem}[Pointwise dichotomy when undoing a change]\label{comp:thm:dicotomia}
Let $\chi:J\to K$ be a $C^\infty$ diffeomorphism, let $b\in J$ and $a=\chi(b)$. Take $\psi=\chi^{-1}$ and consider the reconstruction tasks of Subsection~\ref{subsec:comp-modelo} for \[
 h=(f\circ\chi)\circ\chi^{-1}=f
\] at $a$. Denote its modular cost by $\cm(\chi,b;r)$. Then $\cd=1$ for every $r\ge1$, and exactly one of the following alternatives holds:
\begin{enumerate}[label=\textup{(\alph*)},leftmargin=*]
\item If $\chi^{(m)}(b)=0$ for every $m\ge2$, then $\cm(\chi,b;r)=1$ for every $r\ge1$.
\item If there exists $m\ge2$ with $\chi^{(m)}(b)\ne0$, then \[
 \cm(\chi,b;r)\longrightarrow+\infty\quad(r\to\infty).
\]
\end{enumerate}
\end{theorem}
\begin{proof}
\proofstep{1. Direct task.}
The identity of functions gives $h^{(r)}(a)=f^{(r)}(a)$. One query acquires this datum and checks its availability. Lemma~\ref{comp:lem:necesidad} precludes obtaining it with zero queries. Therefore, $\cd=1$.

\proofstep{2. Case without nonlinear derivatives.}
Suppose that all derivatives of $\chi$ of order at least two vanish at $b$. By \eqref{comp:eq:coef-taylor}, for every $n\ge1$, \[
 B^\chi_{n,k}(b)=0\ (k<n),\qquad B^\chi_{n,n}(b)=c^n,
 \quad c=\chi'(b).
\] Since $(f\circ\chi)\circ\psi=f$, applying \eqref{comp:eq:final}--\eqref{comp:eq:gamma} to $f(y)=(y-a)^k/k!$, for $1\le k\le r$, gives
\[
\sum_{j=k}^{r}B^\psi_{r,j}(a)B^\chi_{j,k}(b)=\delta_{r,k},
\]
where $\delta_{r,k}$ is the Kronecker delta. As $B^\chi_{j,k}(b)=c^j\delta_{j,k}$, we obtain $B^\psi_{r,k}(a)=c^{-r}\delta_{r,k}$. Thus, $V=U=\{r\}$, and Theorem~\ref{comp:thm:costes} gives $\cm=1$.

\proofstep{3. Nonlinear case.}
The coefficient of $g^{(r)}(b)$ in $h^{(r)}(a)$ is \[
 \beta_r=(\psi'(a))^r\ne0.
\] Therefore, every modular reconstruction must recover that intermediate response. Its cost is at least $\#\{k:B^\chi_{r,k}(b)\ne0\}.$ By Lemma~\ref{comp:lem:crecimiento}, this quantity tends to infinity if some higher derivative of $\chi$ does not vanish at $b$. This proves alternative (b). The two alternatives are mutually exclusive and exhaustive.
\end{proof}

\begin{corollary}[Global characterization of bounded modular cost]\label{comp:cor:afin}
For a diffeomorphism $\chi:J\to K$ between open intervals, the following statements are equivalent:
\begin{enumerate}[label=\textup{(\roman*)},leftmargin=*]
\item There exists $M<\infty$ such that $\cm(\chi,b;r)\le M$ for every $b\in J$ and $r\ge1$.
\item $\chi(b)=Ab+B$ for constants $A\ne0$ and $B$.
\item $\cm(\chi,b;r)=1$ for every $b\in J$ and $r\ge1$.
\end{enumerate}
If $\chi$ is not affine, there exists a fixed point $b$ where $\cm(\chi,b;r)\to\infty$.
\end{corollary}
\begin{proof}
If $\chi$ is affine, alternative (a) of Theorem~\ref{comp:thm:dicotomia} holds at every point, so (ii) implies (iii). The implication from (iii) to (i) is immediate.

Suppose (i). If $\chi''(b_0)\ne0$ at some point, alternative (b) of the theorem contradicts the uniform bound. Therefore, $\chi''=0$ on $J$. Since $J$ is an interval, $\chi'$ is constant and $\chi$ is affine. Its slope is nonzero because it is a diffeomorphism. Thus, (i) implies (ii).

Finally, if $\chi$ is not affine, $\chi''=0$ cannot hold throughout $J$. Choose a point with $\chi''(b)\ne0$ and apply the theorem. The chosen point is independent of $r$.
\end{proof}

\begin{remark}[Limits of the statement]\label{comp:rem:limites}
The pointwise conclusion of unit cost does not force $\chi$ to be affine in a neighborhood of $b$: a smooth function may have all its higher derivatives vanish at a point without being affine near it. The global corollary uses every point of an interval. Nor do we claim $\cm=r$ for every nonaffine change or give a uniform growth rate over all changes. For example, $\chi(u)=u+u^3$ at zero has modular cost one at order two and two at order three.
\end{remark}

\subsection{Finite Chains and Matrix Criterion}

Let $\chi_i:J_i\to J_{i-1}$ be known diffeomorphisms, $1\le i\le m$, and fix $t_m\in J_m$. Define recursively \[
 t_{i-1}=\chi_i(t_i),\qquad f_0=f,\qquad f_i=f_{i-1}\circ\chi_i.
\] For a fixed order $r$, consider lower triangular matrices $B_i$ of size $r$, with \[
 (B_i)_{p,k}=B^{\chi_i}_{p,k}(t_i),\qquad 1\le k\le p\le r.
\] The vector of the first $r$ derivatives of $f_i$ at $t_i$ is $B_i$ times the corresponding vector of $f_{i-1}$ at $t_{i-1}$. Consequently, the final coefficients are row $r$ of $B_m\cdots B_1$. This matrix identity has the precedents indicated in Subsection~\ref{subsec:comp-antecedentes}.

We define \[
 S_m=\{r\},\qquad
 S_{i-1}=\bigcup_{p\in S_i}\{k:(B_i)_{p,k}\ne0\},
 \quad i=m,m-1,\ldots,1.
\] The modular reconstruction considered requires the values of the intermediate derivatives with indices in $S_i$, for $1\le i<m$. It is not a minimization over every possible way of deciding which intermediate information to retain.

\begin{proposition}[Final support and modular expansion]\label{comp:prop:cadenas}
For that modular task and the direct task of computing $f_m^{(r)}(t_m)$, \[
 \cm=|S_0|,\qquad
 \cd=\#\{k:(B_m\cdots B_1)_{r,k}\ne0\}.
\] In particular, \[
 1\le\cd\le\cm\le r.
\] Savings are strict exactly when some coordinate of $S_0$ has zero final coefficient.
\end{proposition}
\begin{proof}
By acquiring the data $f^{(k)}(t_0)$, $k\in S_0$, we can successively reconstruct all values required in $S_1,\ldots,S_m$. For $m\ge2$, the list corresponding to $S_1$ contains combinations whose union of supports is $S_0$; Lemma~\ref{comp:lem:necesidad} proves that all these acquisitions are indispensable. For $m=1$, the same argument applies directly to the single requested value.

The direct combination has as its coefficients the indicated row of the product, and the same lemma gives the cost of its support. Every coordinate not belonging to $S_0$ has zero coefficient in that product. Moreover, the nonzero diagonals guarantee that coordinate $r$ always belongs to the final support. All acquired pairs are at $t_0$, which preserves availability of the tasks. The equalities, bounds, and savings criterion follow.
\end{proof}

For a fixed order $r$, increasing only the number of transformations cannot produce a base cost greater than $r$ in this univariate model with storage. The unbounded separation in Corollary~\ref{comp:cor:explog} and Theorem~\ref{comp:thm:dicotomia} is obtained by increasing the derivative order.

\subsection{Joint Interpretation and Limits}\label{subsec:comp-alcance}

Theorem~\ref{comp:thm:costes} compares tasks requiring different information, so savings through simplification do not contradict the optimality in~\ref{thm:chi-optimal}. The dichotomy in~\ref{comp:thm:dicotomia} requires smooth functions and reconstruction of intermediate responses; it makes no claims about time bounds.

\par\medskip
\needspace{8\baselineskip}

\section{Conclusions}\label{sec:conclusiones}

We have established exact representations between computational procedures and query methods with finite budgets. The history-based representation preserves the domain of valid computations that do not exceed the budget, their outputs, records, and counts by order; the reverse construction preserves the same properties. Moreover, we proved that an already acquired response can be reused without modifying the domain or output, although the effective record changes. Staged implementations also admit a representation with exact counts, even when a stage makes no queries.

For a known smooth change of variables, we have determined the minimum cost of reconstructing a derivative of a composite function. Every datum appearing with a nonzero coefficient is necessary, even when adaptive queries at other nodes and orders are allowed. The simulation of a complete procedure preserves its domain, output, and virtual record, and attains the minimum cost by acquiring each active node--order pair in the entire execution only once. Thus, reuse makes it possible to attain simultaneously the minima for total queries and queries by order among faithful simulations.

We then considered successive changes and distinguished reconstruction of intermediate responses from direct computation of the final derivative. We proved that information savings correspond exactly to the active data that disappear through cancellation in the final expression. For the exponential followed by the logarithm, modular reconstruction requires $r$ acquisitions at order $r$, whereas one suffices for direct computation. The criterion extends to finite chains of known changes.

Finally, if a smooth diffeomorphism is composed with its inverse, the direct cost remains equal to one. We proved that the modular cost is one at every order when the higher derivatives of the change vanish at the point under consideration; if any of them does not vanish, that cost tends to infinity as the order increases. Consequently, affine changes are exactly those admitting a uniform modular bound over all points and orders. This conclusion does not require analyticity.

Note that the minima obtained count scalar acquisitions needed to preserve the specified information, rather than internal operations or execution times. Nor are they identified with minimum costs among procedures that produce only the same output. Thus, preservation of domains and records allows us to distinguish computations with identical final values but different information requirements.

\section*{Declaration of generative AI and AI-assisted technologies}

During the preparation of this manuscript, the author used OpenAI's ChatGPT to assist with mathematical development and manuscript revision, and Anthropic's Claude for critical verification of mathematical arguments and formulas. The author reviewed and edited the content and assumes full responsibility for the mathematical results, proofs, references, and final manuscript.

\end{document}